\documentclass[useAMS,usenatbib]{mn2e}

\usepackage{amssymb, amsmath, amsfonts}

\usepackage{graphicx, shadow}

\usepackage{epstopdf}

\usepackage{rotating}
\usepackage{wasysym}
\usepackage{verbatim}
\usepackage{natbib}
\usepackage[draft]{hyperref}

\usepackage{xcolor}

\usepackage{ulem}

\newcommand{\sunrise}{\textsc{Sunrise}}

\newcommand{\hst}{{\sl HST}}
\newcommand{\hstfull}{{\sl Hubble Space Telescope}}
\newcommand{\jwst}{{\sl JWST}}

\newcommand{\illustris}{{Illustris}}

\title[]{Automated Distant Galaxy Merger Classifications from Space Telescope Images using the Illustris Simulation}

\author[Automated Merger Classification Using Illustris]{Gregory F. Snyder$^1$, Vicente Rodriguez-Gomez$^2$, Jennifer M. Lotz$^1$, Paul Torrey$^{3,4}$, \newauthor 
Amanda C.N. Quirk$^{1,5}$, Lars Hernquist$^6$, Mark Vogelsberger$^3$, Peter E. Freeman$^7$ \\
$^1$ Space Telescope Science Institute, 3700 San Martin Dr, Baltimore, MD 21218 \\
$^2$ Department of Physics \& Astronomy, Johns Hopkins University, 3400 N Charles St, Baltimore, MD 21218, USA \\
$^3$ Department of Physics, Kavli Institute for Astrophysics \& Space Research, Massachusetts Institute of Technology, Cambridge, MA, 02139, USA \\
$^4$ Department of Astronomy, University of Florida, 211 Bryant Space Science Center, Gainesville, FL, 32611, USA \\
$^5$ Department of Astronomy \& Astrophysics, UC Santa Cruz, 1156 High St, Santa Cruz, CA 95064 \\
$^6$ Harvard-Smithsonian Center for Astrophysics, 60 Garden St, Cambridge, MA, 02138, USA \\
$^7$ Department of Statistics, Carnegie Mellon University, 5000 Forbes Avenue, Pittsburgh, PA 15213, USA
}
\begin{document}

%%%%%%%%%%%%%%%%%%%%%%%%%%%%%%%%%%%%%%%%%%%%%%%%%%%%%%%%%%%
%%%%%%%%%%%%%%%%%%%%%%%%%%%%%%%%%%%%%%%%%%%%%%%%%%%%%%%%%%%
%%%%%%%%%%%%%%%%%%%%%%%%%%%%%%%%%%%%%%%%%%%%%%%%%%%%%%%%%%%

\maketitle

\begin{abstract}

We present image-based evolution of galaxy mergers from the Illustris cosmological simulation at 12 time-steps over $0.5 < z < 5$. To do so, we created approximately one million synthetic deep Hubble Space Telescope and James Webb Space Telescope images and measured common morphological indicators. Using the merger tree, we assess methods to observationally select mergers with stellar mass ratios as low as 10:1 completing within +/- 250 Myr { of the mock observation}. { We confirm that common one- or two-dimensional statistics select mergers so defined with low purity and completeness, leading to high statistical errors. As an alternative, we train redshift-dependent random forests (RFs) based on 5-10 inputs. Cross-validation shows the RFs yield superior, yet still imperfect, measurements of the late-stage merger fraction, and they select more mergers in bulge-dominated galaxies. When applied to CANDELS morphology catalogs, the RFs estimate a merger rate increasing to at least $z=3$, albeit two times higher than expected by theory. This suggests possible mismatches in the feedback-determined morphologies, but affirms the basic understanding of galaxy merger evolution. The RFs achieve completeness of roughly $70\%$ at $0.5 < z < 3$, and purity increasing from $10\%$ at $z=0.5$ to $60\%$ at $z=3$. At earlier times, the training sets are insufficient, motivating larger simulations and smaller time sampling. By blending large surveys and large simulations, such machine learning techniques offer a promising opportunity to teach us the strengths and weaknesses of inferences about galaxy evolution.
}
\end{abstract}


\section{Introduction} \label{s:intro}

In a Lambda Cold Dark Matter (LCDM) cosmology, galaxies form by the gravity-driven assembly of dark matter halos and accompanying gas. In the early universe, halos of a given mass merge more often than at present owing to a higher matter density, and this prediction has been solidified by cosmological simulations \citep[e.g.,][]{Fakhouri2008,Fakhouri2010,Rodriguez-Gomez2015}.

This process is difficult to observe directly, and only recently have instruments yielded sensitive enough mass tracers to estimate the incidence of very distant galaxy mergers during the epoch of rapid stellar mass growth ($z \gtrsim 2$). Pair fraction measurements coupled with assumptions of a constant observability timescale implied merger rates stopped growing or even declined as the lookback time and fidelity of such surveys increased \citep[e.g.,][]{Ryan2008,Williams2011,Man2016}. However, cosmological simulations suggest that the observability time -- the average duration that a merger could be selected as a specific pair -- must evolve at $z > 1$ \citep{Snyder2017}. Assuming such evolution leads to concordance between measurements and predictions of the massive galaxy merger rate \citep{Ventou2017,Mantha2018}.

Direct measurements of morphological changes caused by galaxy mergers offer complementary information about the merger process. For example, the presence of substantial asymmetry \citep{Conselice2003}, multiple nuclei \citep{Lotz2004}, and very close pairs \citep[e.g.,][]{Lackner2014} have been shown to be promising merger diagnostics.   Human visual classification has been used extensively to select obvious merger candidates \citep[e.g.,][]{Bridge2007,Jogee2009,Darg2010,Kartaltepe2014,Simmons2017}. Moreover, statistical learning techniques have been used to exploit multifaceted information available in high-resolution images measured in both automated and manual ways  \citep[e.g.,][]{Hocking2017,Goulding2018,Ackermann2018}. 

The incidence of late-stage disturbances appears to increase with lookback time to $z \approx 3$ \citep[e.g.,][]{Bluck2012}. However, to judge the effectiveness of a given diagnostic typically requires training against a sample of mergers with its own observational biases, and therefore the resulting conclusions will inherit those limitations. For example, with optical/infrared imaging-based diagnostics, spatially variable dust and stellar populations can confound attempts to measure disturbances induced in the massive stellar components of galaxies \citep{Cibinel2015}. In addition, changes in the general properties of galaxies over time, such as the general increase in star formation (SF) activity, and possible increase in merging activity, can also change how any given diagnostic should be applied in order to yield a consistent physical interpretation \citep[e.g.,][]{Snyder2015a,Abruzzo2018}. Moreover, as mergers evolve, internal structural evolution and star formation can, but doesn't necessarily, wash away information in images about merger initial conditions such as mass ratio \citep{lotz08,lotz10,Lotz2010}. For these reasons, it is not always clear how to compare measurements of late-stage image-based merger diagnostics against theoretical predictions of galaxy assembly processes.

Large cosmological hydrodynamic simulations offer an opportunity to teach us how to more robustly select late-stage mergers.  By combining cosmological context, realistic merger orbit initial conditions, and morphological diversity in large samples, such simulations in principle are the best choice to inform such investigations.  However, they suffer from limitations such as resolution and imperfect feedback physics, which leads to differences in the behavior of some morphological diagnostics \citep[e.g.,][]{Bignone2017}. By contrast, high-resolution zoom simulations tend to overcome the resolution limitation, but may lack the requisite statistical power to sample all merger configurations.  However, they have been proven useful for interpreting image features in order to select distant sources undergoing distinct, more common physical processes \citep{Huertas-Company2018}.

In this paper, we describe how we used a large cosmological hydrodynamic simulation to train image-based diagnostics of distant galaxy mergers. Section~\ref{s:methods} describes the \illustris\ project and our methods to create and characterize synthetic HST and JWST images relevant for major observational programs. Section~\ref{s:mergers} describes and validates optimized merger classification schemes using random forests, an ensemble learning technique. Section~\ref{s:mergerrates} shows how we applied the resulting classification schemes to real HST data from the CANDELS project, and thus how we estimated the observed merger rate in distant massive galaxies to $z=3$.  We discuss the implications of this approach and its challenges in Section~\ref{s:discussion}, and we conclude in Section~\ref{s:conclusion}.

\section{Methods} \label{s:methods}

\subsection{Cosmological Simulation}

To study and improve image-based merger classification techniques, we use the \illustris\ Project simulations of galaxy formation \citep{Vogelsberger2014a,Vogelsberger2014b,Genel2014,Sijacki2015}. \illustris\ simulated a comprehensive phenomenological galaxy formation model \citep{Vogelsberger2013,Torrey2014} throughout a large volume ($106.5$ Mpc$^3$) across cosmic time using the Arepo code \citep{Springel2010}.  The simulation outputs are publicly accessible at www.illustris-project.org/data \citep{Nelson2015}, including catalogs of galaxy properties as well as snapshot output data. 

Achieving decent agreement in global star formation statistics, \illustris\ was among several efforts to successfully simulate a large, diverse galaxy population in reasonable agreement with observations \citep[e.g.][ and others]{Schaye2014, Dubois2014, Khandai2015}.  Using measurements from synthetic optical images at $z=0$, \citet{Snyder2015} showed that \illustris\ achieves a realistic mapping among mass, star formation rate (SFR), kinematics, and morphology. The structures of simulated galaxies have a similar distribution to observed ones in morphology space, and  these morphologies scale appropriately with simulated galaxies' other properties.  Therefore, such simulations meet an important requirement for using them to investigate the physical meaning of measurements of galaxy structure from data.

In this work we restrict ourselves to the simulated galaxies containing enough elements to adequately resolve internal structures, and so we study only simulated sources with a stellar mass above $10^{10} M_{\odot}$ at $z < 6$, and we lower this limit to $10^9 M_{\odot}$ at $z \ge 6$.  Table~\ref{tab:dataset} summarizes the samples.

\begin{figure*}
\begin{center}
\includegraphics[width=6.5in]{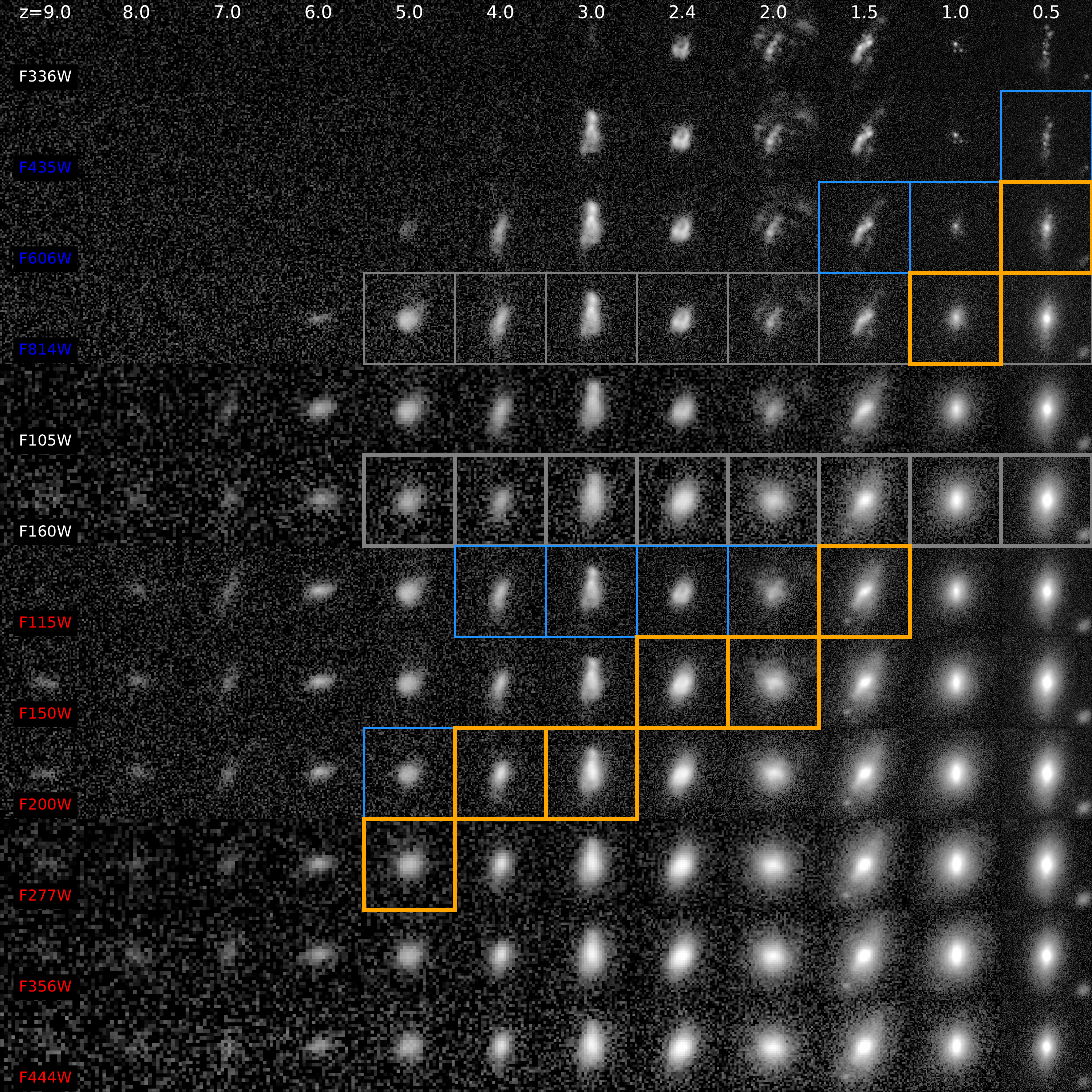}
\caption{\illustris\ synthetic images for one subhalo as it evolves over cosmic time viewed from one direction, in the 12 filters analyzed in this work. This shows our fiducial noise level (``SB25'') and zooms in on a region that is $75(1+z)^{-1}$ ckpc  across. Gray borders show the fixed filter set we study in this paper (relevant for comparison with CANDELS data, for example), while blue and orange borders show the evolving set, which measures similar rest-frame wavelengths and angular resolutions by using \jwst\ filters at $z > 2$. Blue text indicates HST ACS filters, white text indicates HST WFC3 filters, and red text indicates JWST NIRCAM filters. \label{fig:dataset}}
\end{center}
\end{figure*}

\subsection{Pristine Synthetic Images}

We follow \citet{Torrey2015} (hereafter T15) and \citet{Snyder2015} (hereafter S15) to generate broadband images of massive Illustris galaxies in dozens of passbands from 12 snapshots with $z > 0$. Here, we focus on the missions, instruments, and filters listed below and highlighted in Figure~\ref{fig:dataset}, and the galaxy samples listed in Table~\ref{tab:dataset}. We refer the reader to the full details of this procedure in T15 and S15, which we built around the \sunrise \footnote{https://bitbucket.org/lutorm/sunrise } code \citep{jonsson06,jonsson09,Jonsson:2010gpu} for data pipeline compatibility with multiple simulation projects. We host all of our instrument effect-free simulated images at www.illustris-project.org/data.

Differences between the $z=0$ and $z > 0$ sample include a variable physical pixel size, here chosen equal to the JWST's NIRCAM short-wavelength channel pixel size in arcseconds, assuming the \illustris\ cosmology. We arrived at this choice by balancing the desire to sufficiently sample the simulation data ($\Delta R \sim 1.4 (1+z)^{-1}\rm\ kpc$) while limiting data volume to store only pixels that might be relevant for our mock observation analyses. Because this observing mode achieves the best observable spatial resolution for distant galaxy surveys, we use this minimum pixel size for our pristine images so that any realism added will yield mock data with maximum structural information intact, insofar as the simulation supports it. In addition, we applied $(1+z)^{-5}$ cosmological surface brightness dimming in the intrinsic \sunrise\ image units of $W/m/m^2/Sr$.  

We mock-observe simulated galaxies at each of the 12 snapshots in Table~\ref{tab:dataset} from four viewing directions. We define a galaxy to be a subhalo according to the definitions of the halofinding step in the simulation analysis. We do not simulate full lightcone catalogs \citep[c.f.][]{Snyder2017}, and so we neglect distant blends along the line of sight. However, we do include all matter associated with the parent halo of each subhalo we consider. Therefore, our images do include light from external galaxies in ongoing mergers and from overlapping sources in dense regions associated with groups and clusters.

\begin{table*}
\centering
\caption{Dataset Properties.  Unless specified, values correspond to the sample with $M_* > 10^{10.5} M_{\odot}$. $N_{\rm all}$ is the number of \illustris\ subhalos with existing mock images. $N_{\rm massive}$ is the number of subhalos above the more restrictive mass limit of $M_* > 10^{10.5} M_{\odot}$. $N_{\rm images}$ is the number of mock images, where $N_{\rm images}=4*N_{\rm massive}$ because we view each subhalo from four directions. $N_{\rm 10:1}$ and $N_{\rm 4:1}$ show the number of those images that we label as intrinsic mergers based on the criteria in Section~\ref{ss:mergerdef} for merger mass ratio limits of 10 (minor+major) or 4 (major). We define the quantity $\left <M/N\right >$ as the mean number of merger events per object identified as a merger. Many sources identified as a merger will actually experience more than one merger event completing within the desired window. Therefore, we must use this number to compute the correct merger incidence from a sample of binary classifications. }
\label{tab:dataset}
\begin{tabular}{ccc || cccccc}
Snapshot & z & $N_{\rm all} (M_* > 10^{10} M_{\odot}) $ & $N_{\rm massive}$ & $N_{\rm images}$ &  $N_{\rm 10:1}$  &  $N_{\rm 4:1}$ & $\left <M/N\right >_{\rm 10:1}$ & $\left <M/N\right >_{\rm 4:1}$ \\
\hline
% & & $ M_* > 10^{10} M_{\odot}$ & \multicolumn{6}{c}{$ M_* > 10^{10.5} M_{\odot}$}  \\
% \hline
103   &   0.5 &     5486 &     1851 &     7404 &      380 &      176 & 1.2   & 1.3   \\
085   &   1.0 &     4310 &     1417 &     5668 &      524 &      256 & 1.3   & 1.4   \\
075   &   1.5 &     3288 &     1025 &     4100 &      600 &      292 & 1.7   & 1.4   \\
068   &   2.0 &     2233 &      565 &     2260 &      436 &      236 & 1.8   & 1.7   \\
064   &   2.4 &     1569 &      339 &     1356 &      336 &      200 & 1.7   & 1.3   \\
060   &   3.0 &      871 &      137 &      548 &      224 &      132 & 2.2   & 1.6   \\
054   &   4.0 &      233 &       26 &      104 &       52 &       28 & 2.8   & 1.7   \\
049   &   5.0 &     1098 &        3 &       12 &        8 &        8 & 4.0   & 2.5   \\
045   &   6.0 &      361 &        0 &        0 &        0 &        0 & --    & --    \\
041   &   7.0 &       90 &        0 &        0 &        0 &        0 & --    & --    \\
038   &   8.0 &       16 &        0 &        0 &        0 &        0 & --    & --    \\
035   &   9.0 &        3 &        0 &        0 &        0 &        0 & --    & --    \\
\end{tabular}
\end{table*}

\subsection{Simulation of HST and JWST Survey Data}

For this work, we used the sunpy\footnote{https://github.com/ptorrey/sunpy/tree/gfs} Python code to extract and create realistic mock images from our Sunrise calculations \citep{Torrey2015}, the PhotUtils\footnote{v0.3, https://photutils.readthedocs.io/en/stable} package to segment images into separate galaxy objects \citep{Bradley2016}, and custom code to measure image morphology statistics. Our images have pixel scales chosen to reflect the capabilities of HST and JWST to create dithered mosaics achieving spatial resolution commensurate with the telescope aperture and optical design of each instrument, as appropriate, for example the CANDELS data products \citep[][available in MAST: DOI 10.17909/T94S3X ]{Koekemoer2011,Grogin2011}.  Specifically, our mock HST-ACS pixels are 0.03 arcsec, mock HST-WFC3-UV pixels are 0.03 arcsec, mock HST-WFC3-IR pixels are 0.06 arcsec, mock JWST-NIRCAM-Short pixels are 0.032 arcsec, and mock JWST-NIRCAM-Long pixels are 0.064 arcsec.

We convolve each image with an appropriate point-spread function (PSF) modeled with the TinyTim \citep{Krist2011} and WebbPSF \citep{Perrin2014} tools as appropriate.  The resulting data product is then ready to modify as appropriate for any desired effective exposure time.  For our purposes, we create two sets of images with noise added, SB25 and SB27, using a procedure to ensure uniformity across filters and instruments.  We add normally distributed random sky shot noise to each pixel such that the final images achieve a $5\sigma$ limiting surface brightness of 25 and 27 magnitudes per square arcsecond, respectively. Figure~\ref{fig:dataset} presents mock images for each band we consider of the time evolution of a single simulated galaxy.  While these do not correspond to any particular observing strategy, these values were chosen to very roughly correspond to CANDELS-like (SB25) and UDF-like (SB27) depths, respectively.  Our images are somewhat deeper in the bluer ACS filters than was achieved by such surveys, and these choices are simple proxies for possible future JWST observations.  To recreate mock images appropriate for any particular observation, we strongly recommended starting with the noise-free images and deriving one's own noise model to add to each image.

{
In this paper, we use results based on these simple mock images to compare with real HST survey data.  The main justification for this comparison is that the non-parametric morphology measurements we apply are relatively insensitive to the details of the image noise properties, so long as the source is well resolved and bright enough. This has been tested in the literature, recently in \citet{Peth2016}, who finds that all parameters used here are unbiased and reasonably well measured down to a limiting F160W magnitude of approximately 24.5 in CANDELS data.  We trained and applied the RFs only with sources that have an average signal to noise per galaxy pixel of 3 or higher in both filters considered, and we have limited ourselves to fairly massive sources with $M_* > 10^{10.5} M_{\odot}$.  We find that all simulated sources used to train the RFs are brighter than the rough boundary where the morphology measurements break down, an F160W magnitude $\sim 24.0$, and they are often substantially brighter (F160W $ < 22$). While fainter features, such as tidal tails, would be strongly affected by surface brightness limitations or the spatial correlations in true HST noise, in this study, we are not using morphology measurements on simulated or real data that are near their intended surface brightness limit.
}
%\begin{figure}
%\begin{center}
%\includegraphics[width=3.5in]{fig2.pdf}
%\caption{(draft) Approximate spatial resolution of \illustris\ synthetic images versus time, compared to estimated resolving power of example \hst\ and \jwst\ instrument and filter combinations. In our synthetic images, all original images have a pixel size equal to the NIRCAM-short instrument pixel scale ($0.0165$ arcsec) for the \illustris\ cosmology. This choice samples all PSF cores used in this work with at least two pixels in each dimension, and adequately samples the intrinsic \illustris\ spatial resolution given conservatively by $2\times$ the minimum gravitational softening. At all cosmic times, this scale is similarly or better resolved than the corresponding survey capabilities. \label{fig:scales}}
%\end{center}
%\end{figure}

\subsection{Morphology Statistics} \label{ss:morphstats}

We measure non-parametric morphology statistics designed to characterize the light profile shape as well as the presence of multiple nuclei or disturbances. These include the Concentration, Asymmetry statistics \citep[e.g.][]{Conselice2003}, the Gini, $M_{20}$ statistics \citep[e.g.][]{Lotz2004}, and the M, I, and D statistics of \citet{Freeman2013}.  Specifically, we follow \citet{Snyder2015a} and \citet{Snyder2015} to analyze the simulated images and refer readers to the specific choices outlined therein. To simplify discussion, we rotate the $G$-$M_{20}$ space to define two composite diagnostics, a ``bulge statistic'' $F$ and a ``merger statistic'' $S$:
\begin{equation} \label{eq:fgm20}
F(G,M_{20}) =
\begin{cases}
\left | F\right | & G \ge 0.14 M_{20} + 0.778\\
-\left | F\right | & G < 0.14 M_{20} + 0.778,
\end{cases} 
\end{equation}
where $\left | F \right | = \left | -0.693 M_{20}  + 4.95 G - 3.85 \right |$ \citep{Snyder2015}.
\begin{equation} \label{eq:sgm20}
S(G,M_{20}) =
\begin{cases}
\left | S\right | & G \ge -0.14 M_{20} + 0.33\\
-\left | S\right | & G < -0.14 M_{20} + 0.33,
\end{cases} 
\end{equation} 
where $\left | S \right | = \left | 0.139 M_{20}  +  0.990 G - 0.327 \right |$ \citep{Snyder2015a}.

%% TP tp_masshistory_snapshot_075_sh61401_cam1  tp_masshistory_snapshot_075_sh44389_cam0
%% FP fp_masshistory_snapshot_075_sh80114_cam1  fp_masshistory_snapshot_075_sh32127_cam2
%% TN tn_masshistory_snapshot_075_sh75664_cam1  tn_masshistory_snapshot_075_sh69862_cam1
%% FN fn_masshistory_snapshot_075_sh86695_cam2  fn_masshistory_snapshot_075_sh120495_cam3

\begin{figure*}
\begin{center}
\includegraphics[width=3.1in]{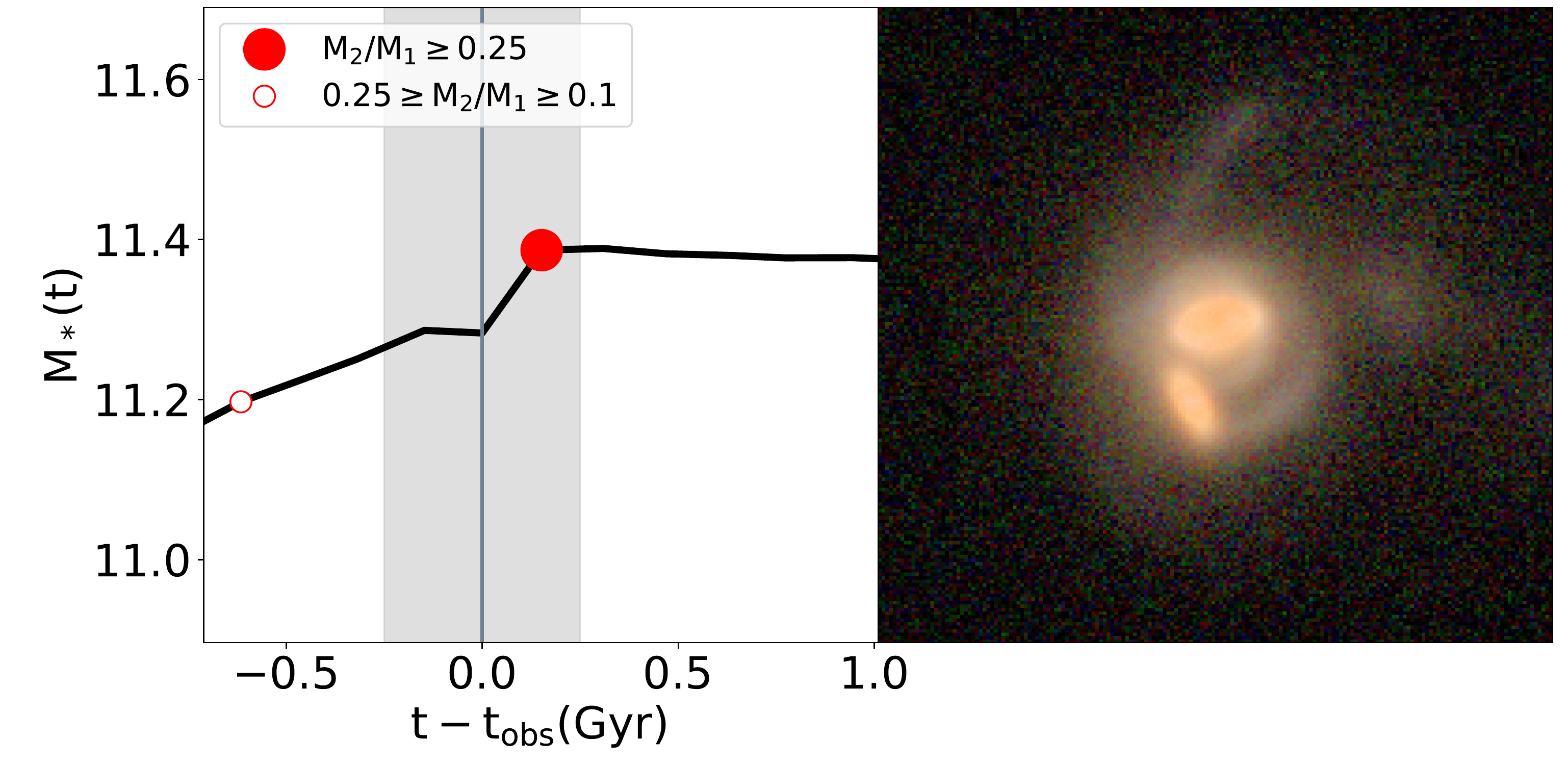}
\includegraphics[width=3.1in]{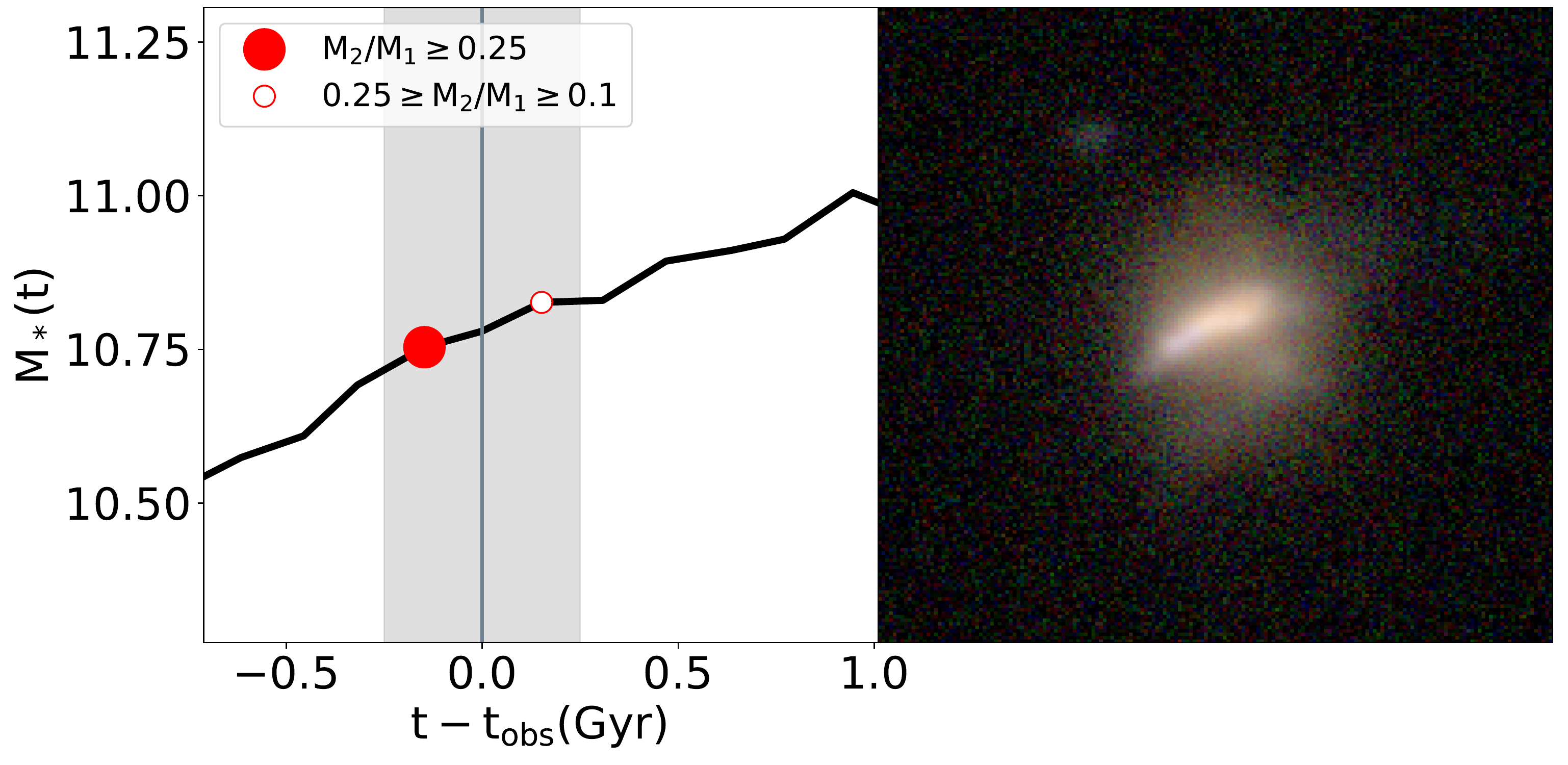}

\includegraphics[width=3.1in]{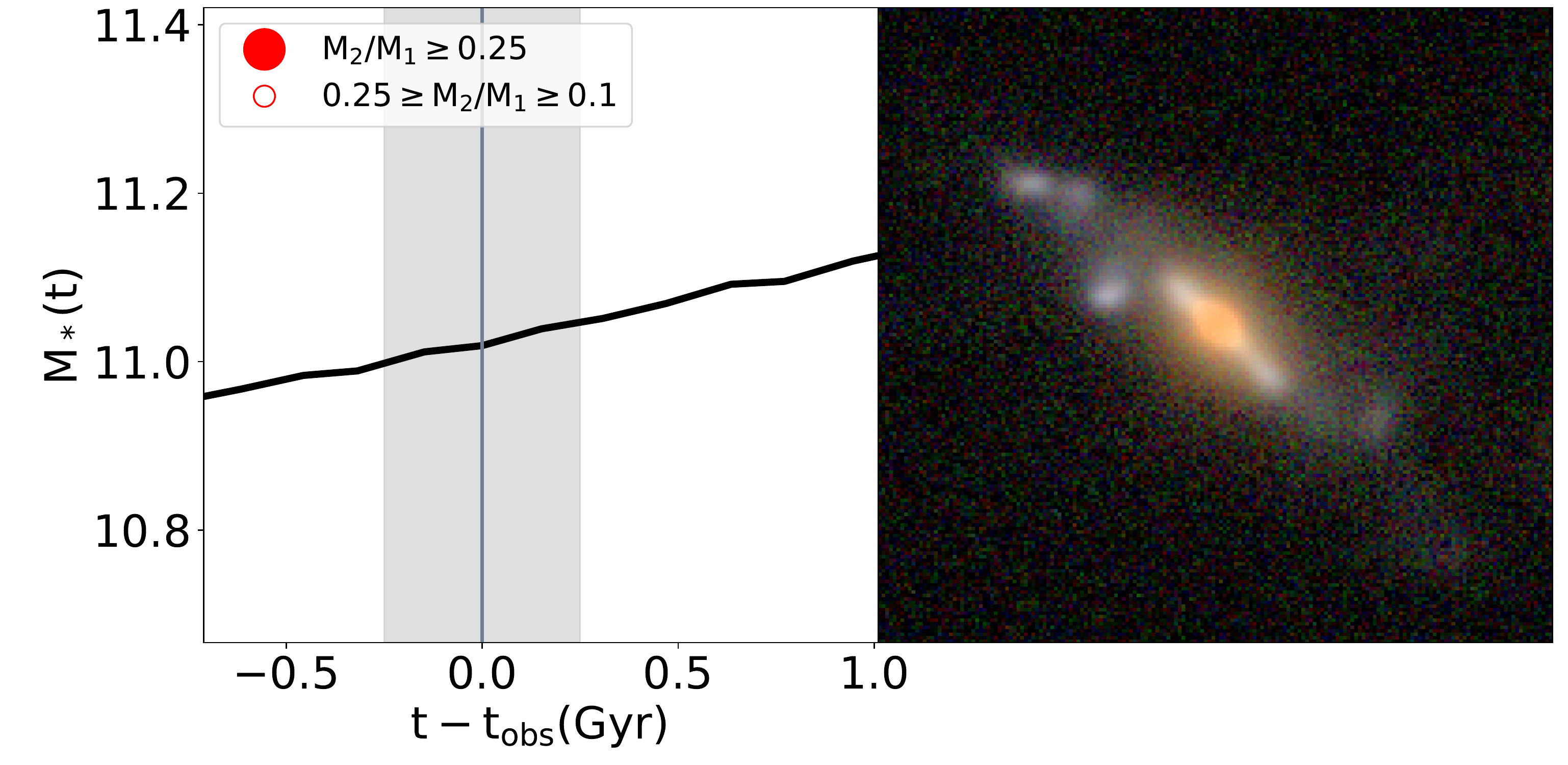}
\includegraphics[width=3.1in]{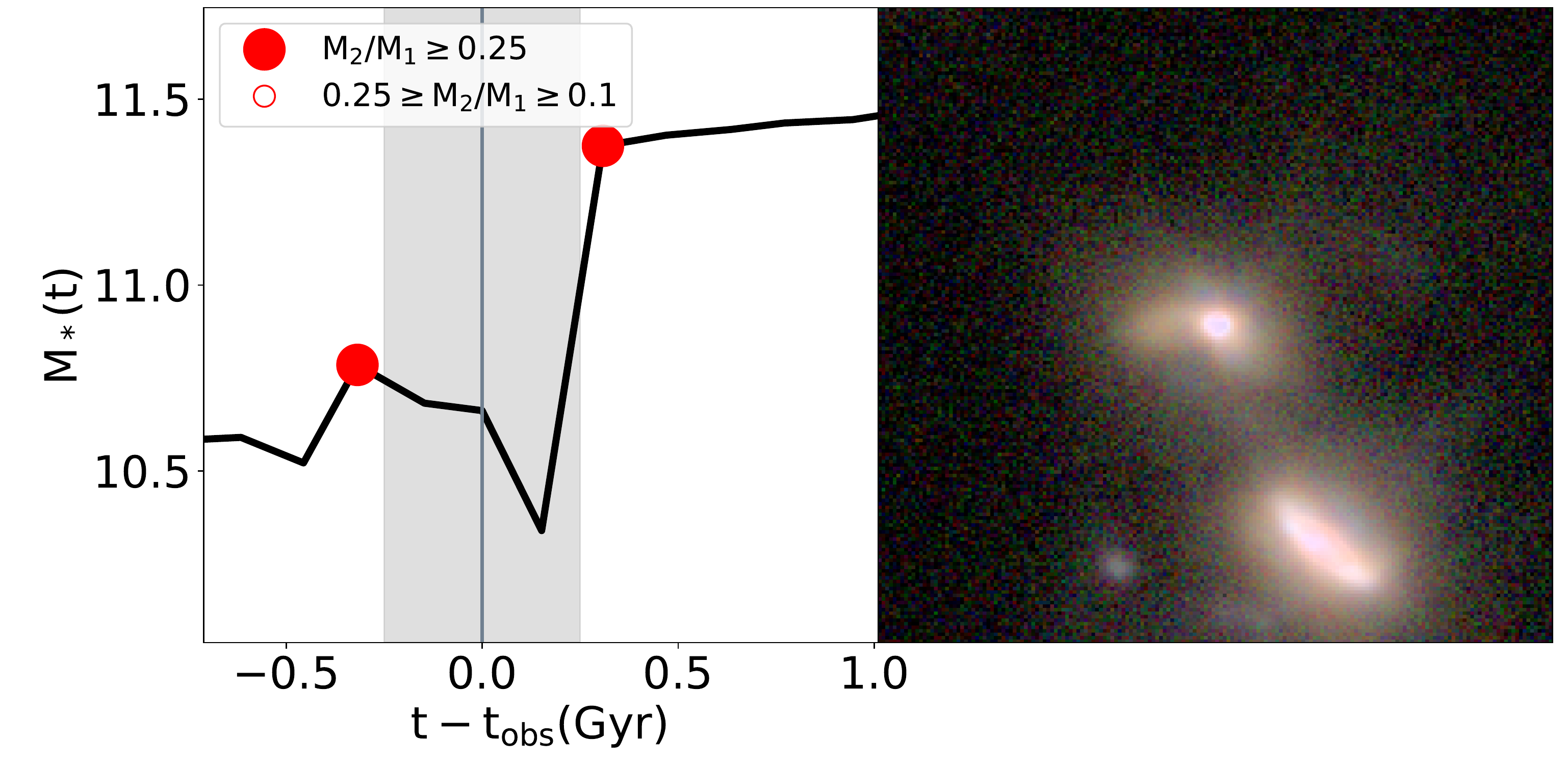}

\includegraphics[width=3.1in]{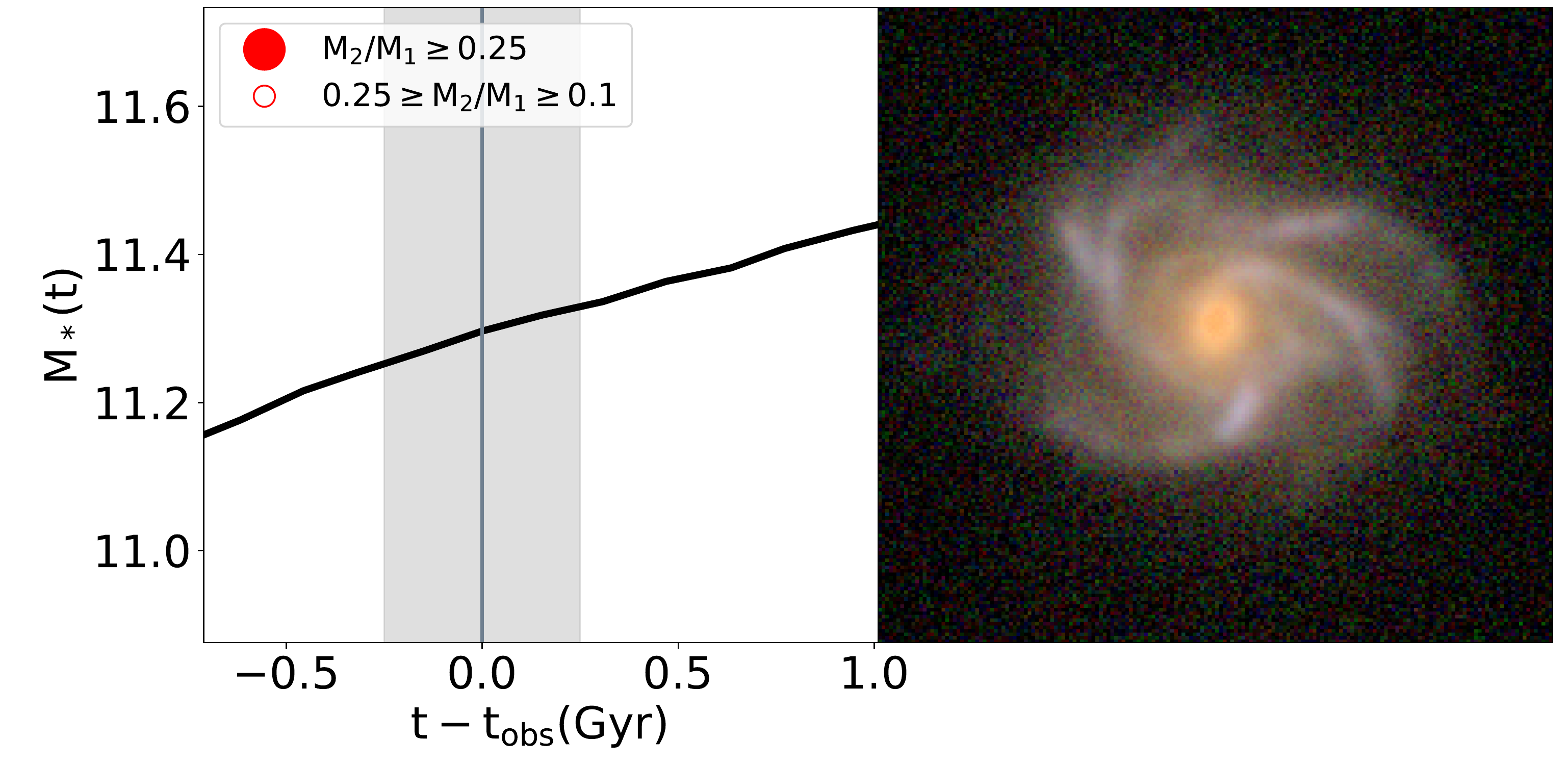}
\includegraphics[width=3.1in]{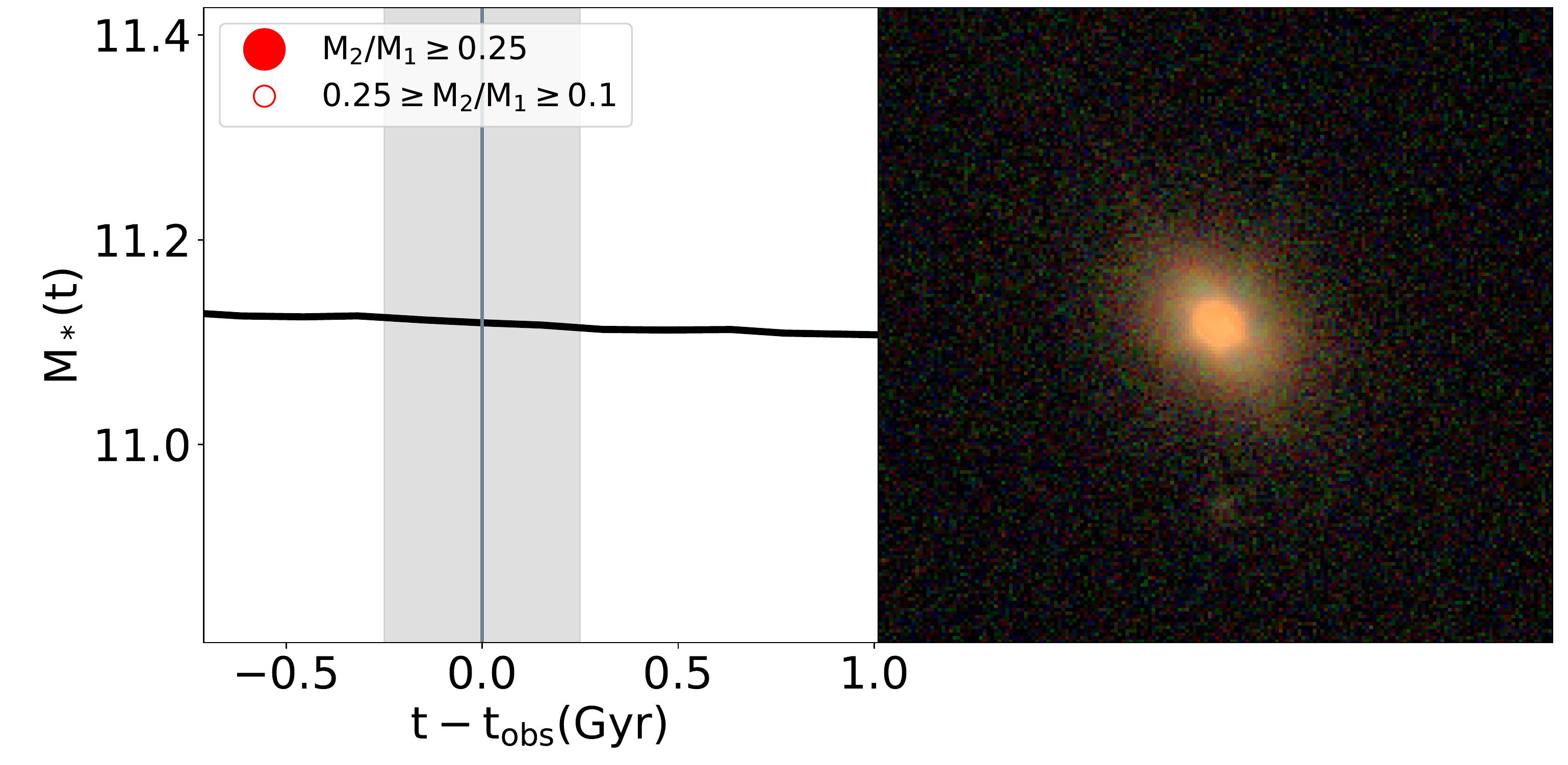}

\includegraphics[width=3.1in]{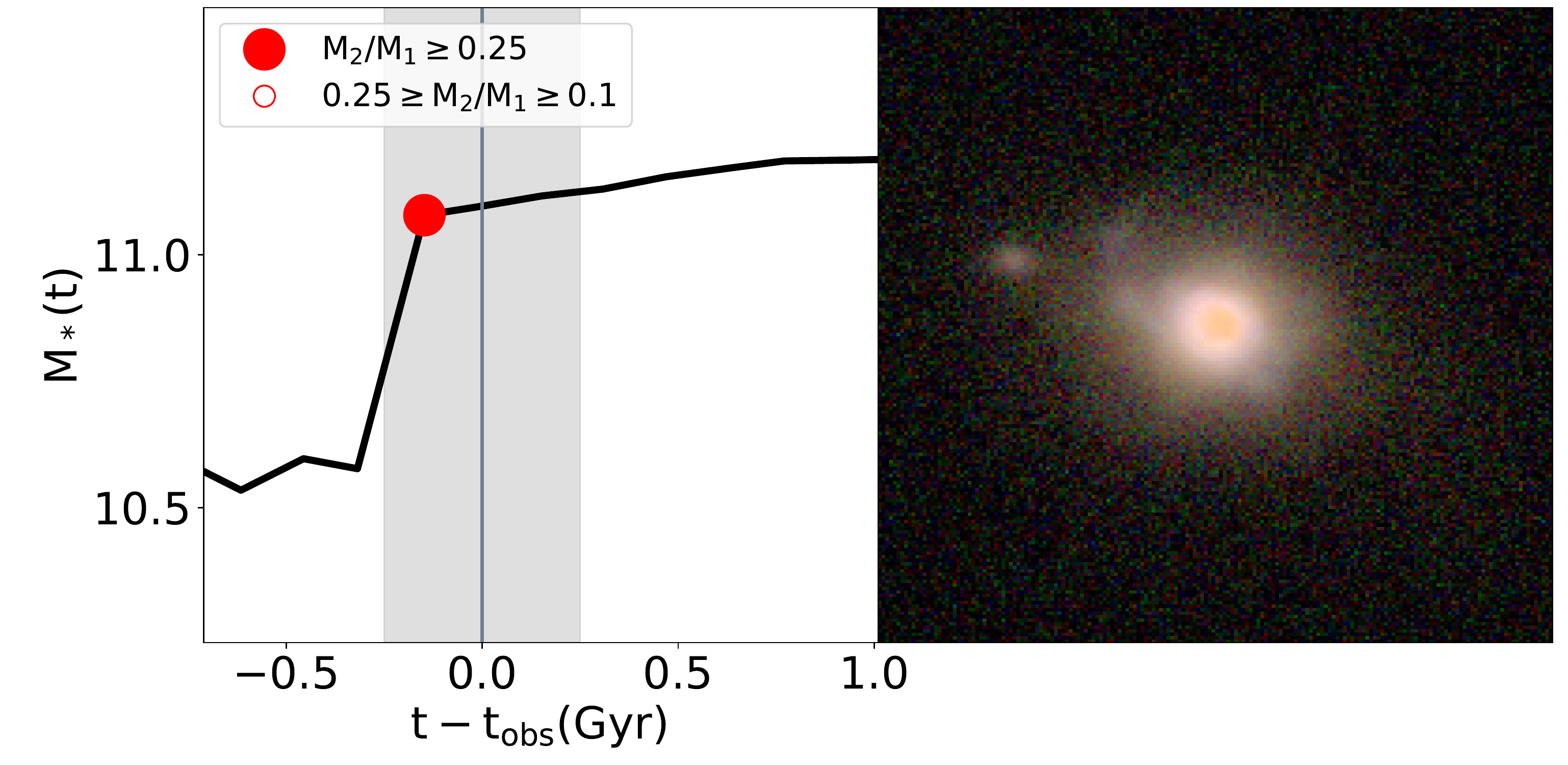}
\includegraphics[width=3.1in]{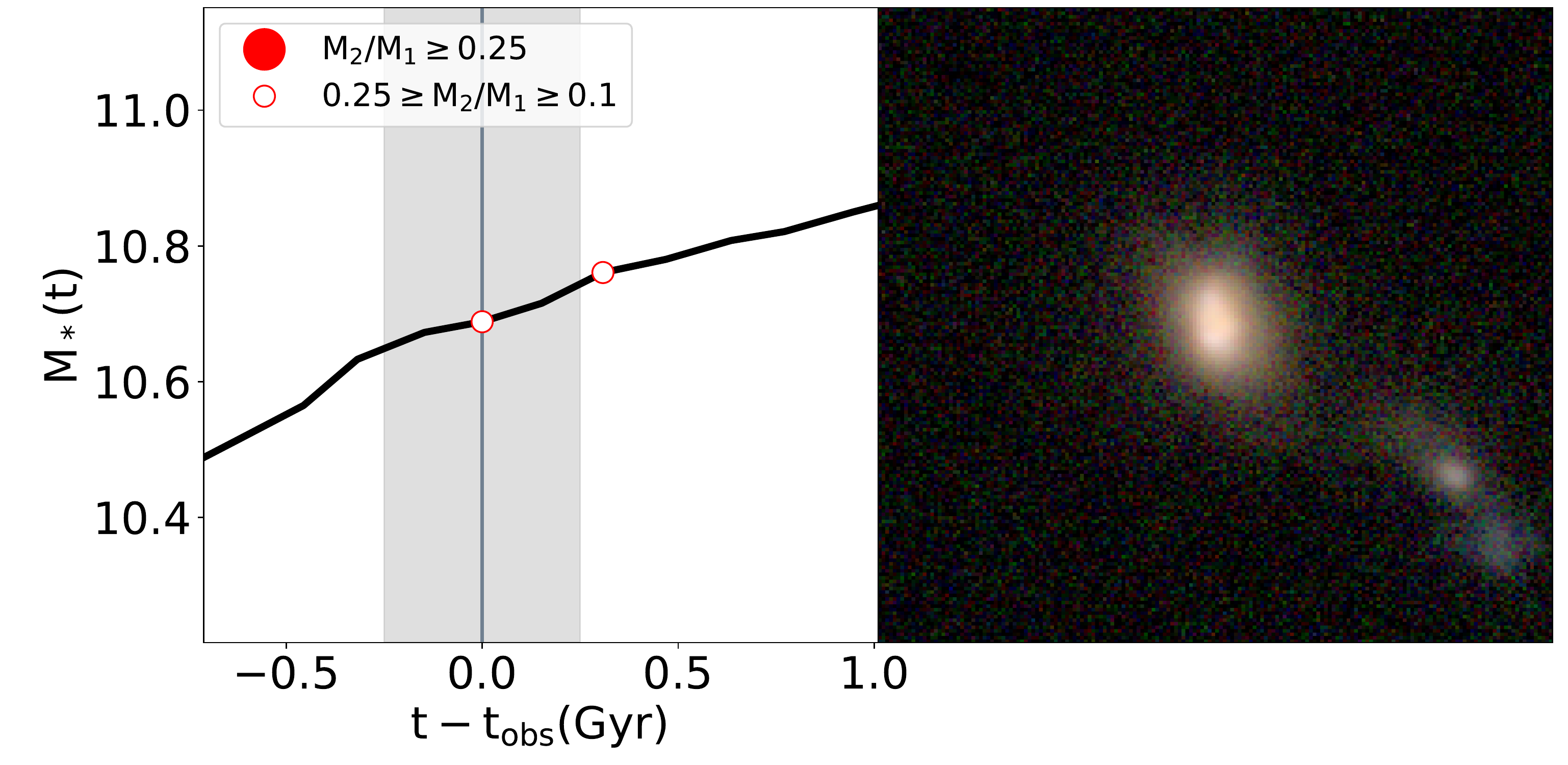}

\caption{Eight example sources in our mock image training sample at $z=1.5$, demonstrating the variety of systems and merger states manifest in \illustris. For each source, the left panel shows the time-evolution of the stellar mass of the main progenitor, with red circles indicating major (filled) and minor (open) mergers as defined in Section~\ref{ss:mergerdef}. For clarity, we consider only the last and next mergers in each category. If such a merger occurs within the gray shaded region of $\pm 250$ Myr, we label the source as a merger event (Section~\ref{ss:mergerdef}) for the purpose of investigating classifier performance. We also show three-color composite images of each source rendered from mock JWST NIRCAM F115W, F150W, and F200W images.  \label{fig:merger_examples}}
\end{center}
\end{figure*}

\subsection{Intrinsic Merger Definition}  \label{ss:mergerdef}

\citet{Rodriguez-Gomez2015} (R-G15) analyzed \illustris\ to define merger trees using the SubLink code. From these merger trees, we defined catalogs of merger events relative to each galaxy measured in our mock image program (Table~\ref{tab:dataset}).  These catalogs include a range of different merger definitions, allowing us to choose various mass ratios and timescales with which to define a sample of true intrinsic mergers.  

Figure~\ref{fig:merger_examples} presents example images and mass growth histories to demonstrate the merger selections we use in this paper.  We use limiting stellar mass ratios of 10:1 or 4:1, events commonly labeled minor+major and major mergers respectively, as defined by R-G15: the mass ratio at the time the secondary achieved its highest mass prior to the merger.  We focus on one of many possible time windows to define a sample of intrinsic mergers nearby in time to the mock-observed galaxies: a merger completes within a 0.5 Gyr window centered on the time of observation ($| t_{\rm merge}-t_{\rm obs} | \leq 0.25$ Gyr). We have explored several other definitions, such as shorter and longer windows, as well as a $1.0$ Gyr window centered at $t_{\rm obs} + 0.25$ Gyr, all of which achieve similar performance and limitations in the classification tests of Section~\ref{s:mergers}, and so we omit them from this paper for clarity and brevity. 

A unique aspect of our definition is the relatively broad set of objects used to label the intrinsic merger sample, by using a relatively wide time window ($500$ Myr) and including mergers with mass ratios up to 10.  One reason for this was that we required a large enough training set to carry out the calculations in Section~\ref{s:mergers}.  However, a benefit of this choice is that we might be able to identify merger signatures more subtle or long-lived than the ones conventionally used. For example, \citet{lotz08} showed that classical image-based merger definitions have observability times of only 100-200 Myr.  Therefore the broader definition we have chosen could have important consequences for the relative performance of different merger indicators.

Regardless of which mergers we choose as our intrinsic sample, there will always be situations that make it more or less difficult for us to identify them accurately. As Figure~\ref{fig:merger_examples} shows, it is possible for an image to appear very much like an ongoing merger (for example, a close pair or multiple nuclei) even though a merger event completes outside of these two selections and therefore that object would not be considered a merger in this work. This false positive mode is an important failure mode for our attempts to define ideal image-based merger criteria in Section~\ref{s:mergers}. It may be possible to derive a more informative merger classification scheme by allowing for more continuous merger event definitions, for example including mergers that complete more than 0.25 Gyr in the future. This type of selection would then more closely mimic the one used to assess samples of merging pairs \citep[e.g.][]{Snyder2017}.

Moreover, it is possible for images to appear rather isolated very quickly after the merger completes or prior to a merger with a companion too distant to influence the image. These mergers may be very difficult to identify using image morphology criteria alone, leading to confusion between mergers and nonmergers. On the other hand, these galaxies may exhibit subtle cues that visual or standard classifications could miss, and so new techniques to automatically classify them may bear fruit.

Table~\ref{tab:sim_morph_stats} shows example entries of the \illustris\ galaxy properties, merger states, and morphology measurements we study in this paper. We include the full table as supplementary material.

\begin{table*}
\centering
\caption{Simulation morphology catalog, example entries, first 16 columns and 42 rows. We include the full catalog as online-only supplementary material.  Entries with ``nan'' result from unsuccessful morphology measurements, while ``None'' indicates a source that was not selected as part of the random forest training sample in Section~\ref{s:mergers}.}
\label{tab:sim_morph_stats}
\begin{tabular}{cc cccc ccccc ccccc}
Snap & Sub ID & $log_{10} \frac{M_*}{M_{\odot}}$ & Cam & Merger & $P_{RF}$ & $A_I$ & $C_I$ & $G_I$ & $M_{20,I}$ & $D_I$ & $A_H$ & $C_H$ & $G_H$ & $M_{20,H}$ & $D_H$  \\
103    &     81534  & 11.67  &  3  &False   &None   &  0.05 &  3.98 &  0.63 & -1.47 & -1.37 &   nan &   nan &   nan &   nan &     nan \\ 
103    &     95704  & 11.66  &  3  &False   &0.018  & -0.03 &  4.22 &  0.62 & -2.32 & -2.02 &  0.01 &  4.05 &  0.60 & -2.25 & -1.6302 \\ 
103    &    107679  & 11.64  &  2  &False   &0.055  &  0.16 &  2.55 &  0.53 & -1.24 & -0.25 &  0.09 &  2.88 &  0.54 & -1.72 & -1.3492 \\ 
103    &     67492  & 11.63  &  3  &False   &0.017  & -0.01 &  3.27 &  0.57 & -1.82 & -1.67 &  0.01 &  3.27 &  0.57 & -1.85 & -1.7325 \\ 
103    &    127701  & 11.62  &  2  &False   &0.018  & -0.03 &  4.75 &  0.64 & -2.51 & -2.23 &  0.01 &  4.47 &  0.62 & -2.42 & -1.6726 \\ 
103    &     86431  & 11.61  &  2  &False   &None   &   nan &   nan &   nan &   nan &   nan &  0.36 &  3.54 &  0.58 & -0.69 & -0.4011 \\ 
085    &     46869  & 11.55  &  3  &False   &0.04   & -0.02 &  3.39 &  0.60 & -1.86 & -1.63 &  0.02 &  3.20 &  0.57 & -1.77 & -1.6265 \\ 
085    &     91678  & 11.55  &  3  &True    &0.387  &  0.06 &  3.37 &  0.56 & -2.03 & -1.70 &  0.05 &  3.58 &  0.56 & -2.16 & -1.9818 \\ 
085    &     44494  & 11.54  &  1  &False   &0.04   &  0.02 &  3.24 &  0.52 & -1.97 & -1.61 &  0.05 &  3.48 &  0.53 & -2.11 & -1.7076 \\ 
085    &     82223  & 11.54  &  3  &False   &0.252  &  0.10 &  4.24 &  0.64 & -1.93 & -1.09 &  0.07 &  3.86 &  0.61 & -1.99 & -1.1703 \\ 
085    &     56730  & 11.54  &  2  &True    &0.155  & -0.01 &  3.45 &  0.58 & -1.98 & -1.37 &  0.04 &  3.59 &  0.56 & -2.11 & -1.7937 \\ 
085    &     69745  & 11.53  &  2  &True    &0.68   &  0.37 &  2.44 &  0.62 & -1.50 & -0.61 &  0.24 &  4.02 &  0.62 & -2.02 & -1.1677 \\ 
075    &     28283  & 11.46  &  3  &False   &0.089  &  0.03 &  2.57 &  0.53 & -1.51 & -0.90 &  0.02 &  2.95 &  0.55 & -1.79 & -1.4321 \\ 
075    &     46424  & 11.46  &  0  &False   &0.123  &  0.37 &  2.21 &  0.46 & -0.76 & -0.41 &  0.19 &  3.01 &  0.52 & -1.71 & -1.1012 \\ 
075    &     45512  & 11.45  &  2  &False   &0.258  &  0.34 &  2.57 &  0.49 & -1.18 & -0.63 &  0.12 &  3.00 &  0.51 & -1.75 & -1.7291 \\ 
075    &     24678  & 11.44  &  3  &False   &0.449  &  0.38 &  2.68 &  0.59 & -0.71 & -0.88 &  0.24 &  3.81 &  0.57 & -1.60 & -1.0966 \\ 
075    &     41018  & 11.44  &  3  &True    &None   &  0.09 &  3.03 &  0.52 & -1.67 &  0.03 &  0.05 &  3.43 &  0.54 & -2.01 & -1.3146 \\ 
075    &     68391  & 11.44  &  2  &False   &None   & -0.18 &   nan &   nan &   nan &   nan & -0.02 &  4.26 &  0.60 & -2.12 & -1.0611 \\ 
068    &     24677  & 11.28  &  2  &False   &0.164  &  0.07 &  3.36 &  0.52 & -0.96 & -0.05 &  0.06 &  3.52 &  0.51 & -2.00 & -1.2420 \\ 
068    &     31101  & 11.27  &  3  &False   &0.273  &  0.22 &  2.32 &  0.50 & -1.25 & -0.64 &  0.08 &  2.66 &  0.52 & -1.57 & -1.1113 \\ 
068    &     22649  & 11.26  &  3  &False   &0.24   &  0.35 &  2.27 &  0.53 & -1.17 & -0.52 &  0.15 &  2.86 &  0.51 & -1.74 & -1.2378 \\ 
068    &     27823  & 11.26  &  1  &False   &0.39   &  0.64 &  2.54 &  0.63 & -1.09 & -0.43 &  0.30 &  3.61 &  0.56 & -1.75 & -0.9295 \\ 
068    &     16521  & 11.26  &  0  &False   &0.263  &  0.15 &  2.26 &  0.48 & -1.08 & -0.98 &  0.07 &  3.07 &  0.61 & -1.39 & -1.3815 \\ 
068    &     21895  & 11.25  &  1  &False   &None   &   nan &   nan &   nan &   nan &   nan & -0.04 &  4.37 &  0.64 & -2.35 & -1.7616 \\ 
064    &     23563  & 11.17  &  1  &False   &None   &   nan &   nan &   nan &   nan &   nan & -0.04 &  3.11 &  0.58 & -1.81 & -1.5585 \\ 
064    &      2709  & 11.16  &  3  &True    &0.454  &  0.46 &  3.14 &  0.54 & -1.52 & -0.52 &  0.27 &  3.29 &  0.53 & -1.69 & -0.8457 \\ 
064    &     15173  & 11.16  &  0  &False   &0.147  &  0.13 &  2.39 &  0.44 & -1.29 & -0.58 &  0.06 &  2.93 &  0.52 & -1.80 & -1.2799 \\ 
064    &      2710  & 11.15  &  0  &False   &0.339  &  0.18 &  3.38 &  0.56 & -1.93 & -1.92 &  0.05 &  3.42 &  0.59 & -1.96 & -1.3107 \\ 
064    &       962  & 11.15  &  2  &True    &0.779  &  0.31 &  3.91 &  0.65 & -0.91 & -0.87 &  0.21 &  4.12 &  0.62 & -1.70 & -1.1554 \\ 
064    &     13206  & 11.14  &  3  &False   &0.171  &  0.07 &  2.47 &  0.44 & -1.11 & -1.00 &  0.03 &  3.41 &  0.53 & -1.93 & -1.4747 \\ 
060    &      7525  & 10.93  &  3  &True    &0.307  &  0.26 &  2.97 &  0.59 & -1.31 & -0.84 &  0.14 &  3.39 &  0.59 & -1.83 & -0.9349 \\ 
060    &     12542  & 10.92  &  0  &True    &0.229  &  0.13 &  3.06 &  0.57 & -1.42 & -0.35 &  0.03 &  3.01 &  0.56 & -1.66 & -0.8412 \\ 
060    &      7679  & 10.92  &  1  &False   &0.229  &  0.07 &  3.19 &  0.60 & -1.59 & -0.74 & -0.01 &  3.21 &  0.57 & -1.73 & -1.2706 \\ 
060    &     16998  & 10.91  &  1  &False   &None   & -0.15 &  2.63 &  0.49 & -1.46 &  0.24 & -0.00 &  2.55 &  0.50 & -1.65 & -0.8641 \\ 
060    &     15836  & 10.91  &  2  &False   &None   & -0.10 &   nan &   nan &   nan &   nan & -0.08 &  3.25 &  0.60 & -1.91 & -1.3136 \\ 
060    &     11090  & 10.91  &  1  &True    &0.572  &  0.31 &  2.52 &  0.52 & -1.12 & -0.77 &  0.09 &  2.88 &  0.53 & -1.49 & -0.9723 \\ 
054    &      5005  & 10.52  &  0  &True    &0.788  &  0.27 &  2.80 &  0.50 & -1.38 & -0.88 &  0.12 &  2.68 &  0.49 & -1.68 & -1.0138 \\ 
054    &      7442  & 10.48  &  1  &True    &None   &  0.31 &  2.61 &  0.48 & -1.31 & -0.37 &  0.09 &  2.33 &  0.46 & -1.36 & -0.4802 \\ 
054    &      3972  & 10.45  &  0  &False   &None   &  0.00 &  2.78 &  0.54 & -0.95 & -0.16 & -0.06 &  2.78 &  0.48 & -1.27 &  0.2607 \\ 
054    &      9923  & 10.44  &  2  &False   &None   &  0.21 &  2.20 &  0.53 & -1.06 & -0.27 &  0.08 &  2.54 &  0.50 & -1.33 & -0.6043 \\ 
054    &     10296  & 10.42  &  3  &False   &None   &  0.26 &  1.86 &  0.45 & -0.92 & -0.54 &  0.09 &  2.20 &  0.46 & -1.22 & -1.1872 \\ 
054    &      7959  & 10.42  &  0  &False   &None   &  0.28 &  3.08 &  0.57 & -1.59 & -0.05 &  0.05 &  2.93 &  0.54 & -1.64 & -0.2833 \\ 
\end{tabular}
\end{table*}

\begin{figure*}
\begin{center}
\includegraphics[width=2.25in]{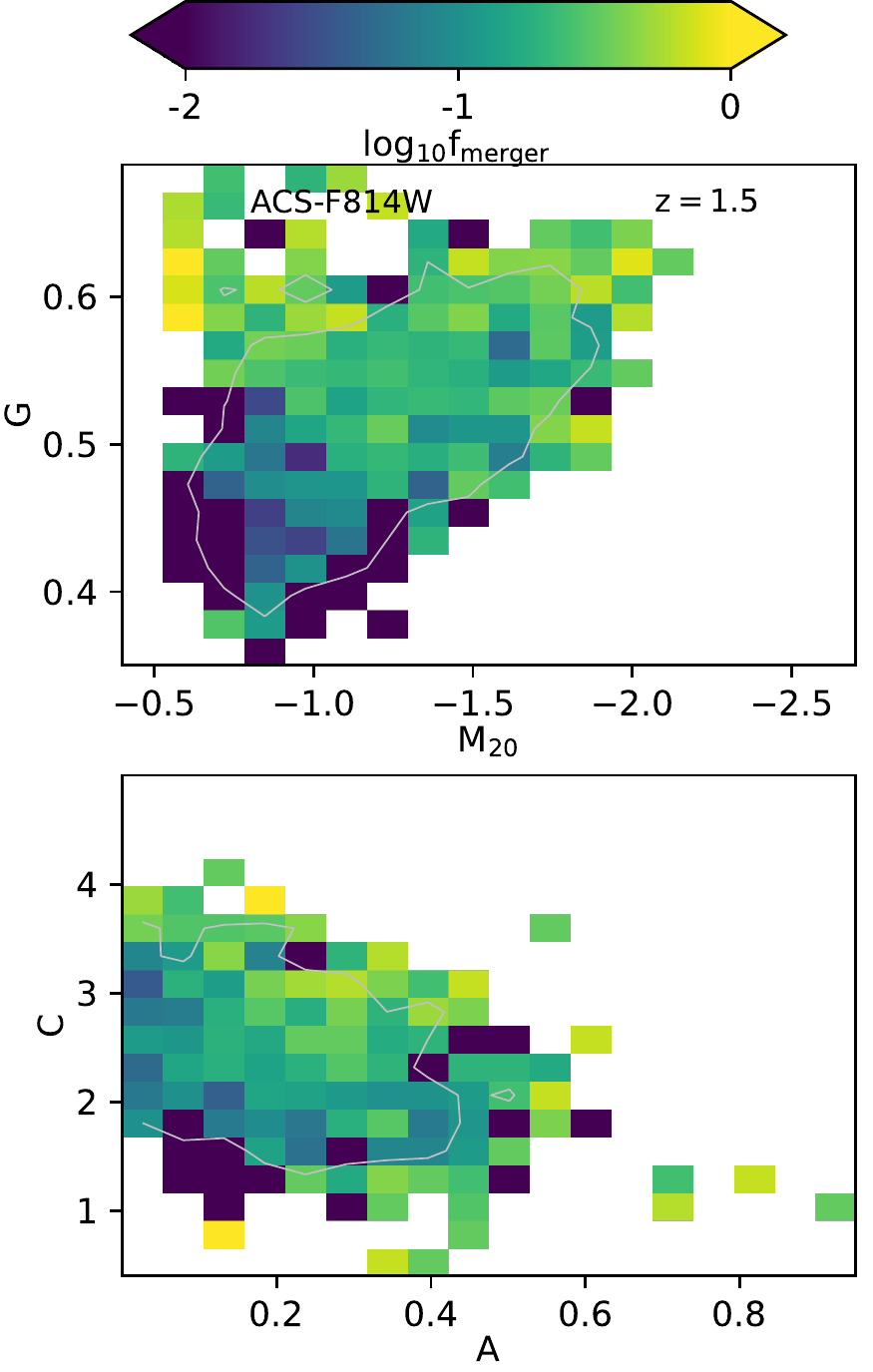}\includegraphics[width=2.25in]{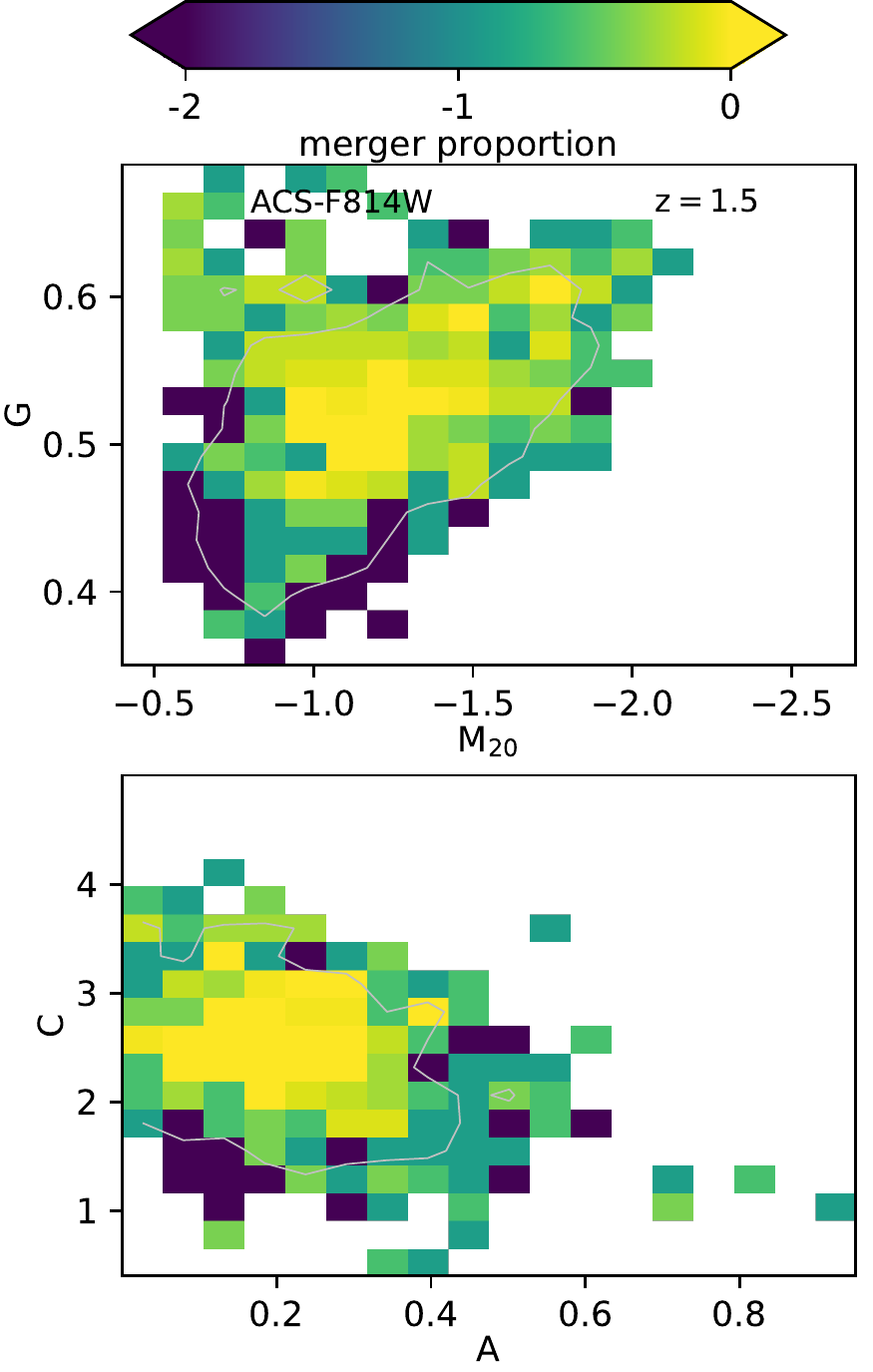}\includegraphics[width=2.25in]{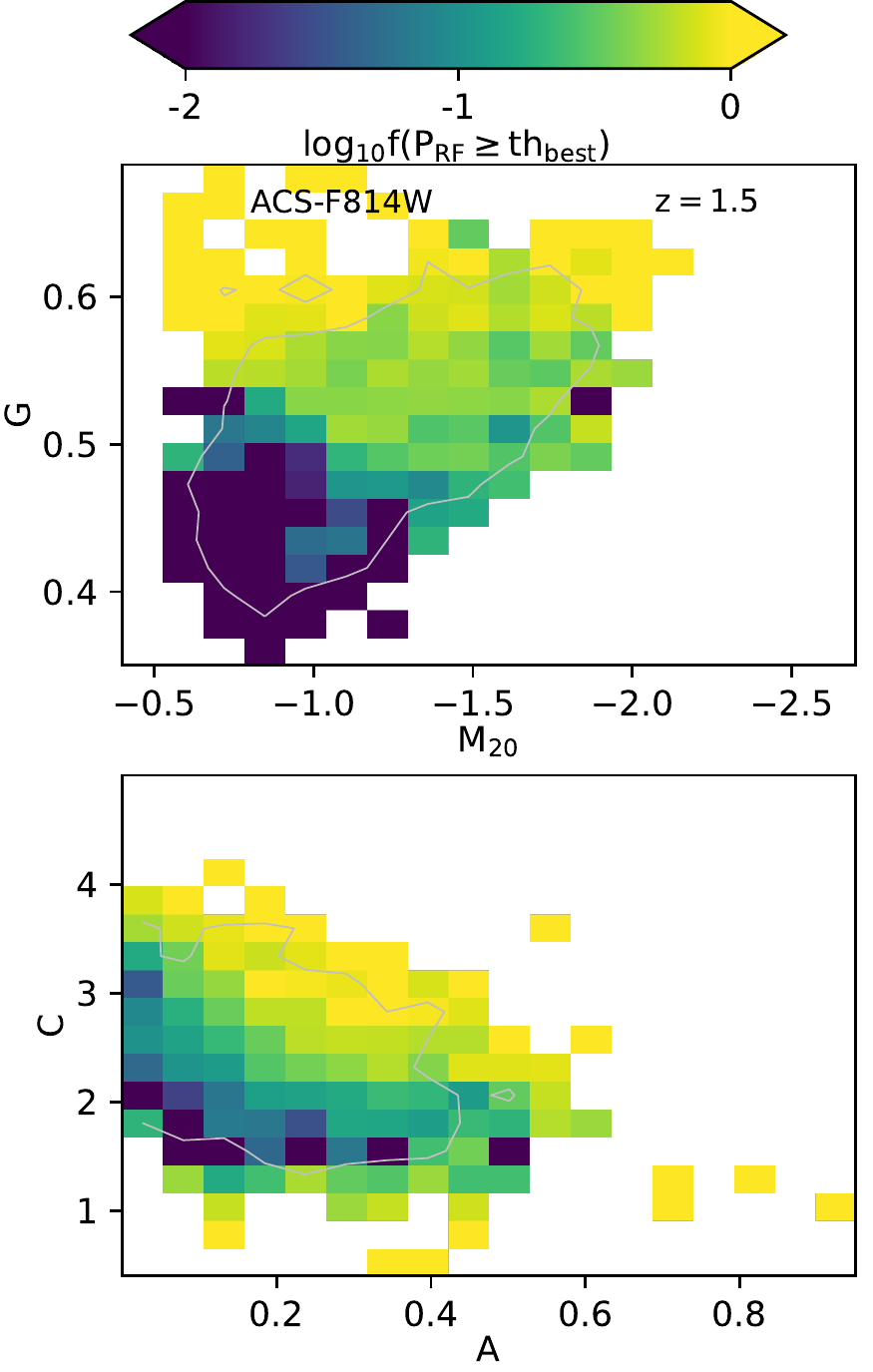}
\caption{\illustris\ morphology diagrams at $z=1.5$ for $\approx 4000$ sources ($\approx 600$ mergers), color-coded by merger fraction (left), proportion of all mergers (center), and our random forest-classified merger fraction (right).  The simplest morphological merger classifications select the upper left of the $G$-$M_{20}$ panel and right side of the $C$-$A$, regions which do contain a high fraction of mergers (left). However, most mergers do not have these unusual morphologies and therefore occupy the same region of parameter space as average non-merging galaxies (center). Therefore, to select superior merger samples, we seek to define new classification techniques (e.g., right) that better trace the distribution of mergers in this space. This is seen as the lighter shades of the right column spanning more of parameter space, capturing more mergers that do not occupy the standard regions used to classify mergers. \label{fig:classical} }
\end{center}
\end{figure*}

\subsection{Low-dimensional Merger Classifications}

With an intrinsic merger definition and a suite of morphology measurements in hand for the galaxy samples in Table~\ref{tab:dataset}, we now quantify the performance of several classical merger diagnostics.  

The left and middle columns of Figure~\ref{fig:classical} show where the set of intrinsic mergers fall within two-dimensional diagnostic diagrams sometimes used to isolate mergers, the spaces of $G$ versus $M_{20}$ and $C$ versus $A$.  In these plots we separate the space into discrete bins, where each bin contains objects with similar morphology measurements, and compute simple quantities about the merger populations.  In the left column, we show the fraction of sources in each bin that are mergers according to our definition.  In the center column, we show a value indicating what fraction of all the intrinsic mergers falls in each bin, where this fraction has been normalized to peak at a value of 1.0.  

These two ways of displaying the merger population in Figure~\ref{fig:classical} correspond to quantifying the purity and completeness of mergers in each bin, respectively.  For example, the left column can be thought of as showing that some regions of the space of $G$-$M_{20}$ and $C$-$A$ can select samples of mergers with relatively high purity, such as objects with high $A$ and objects with high $G$ and $M_{20}$.  We find a result similar to the one found by \citet{Bignone2017} that sources with high Asymmetry have lower purity ($\lesssim 50\%$) than thought to exist in real sources, perhaps owing to the fact that \illustris\ galaxies have higher levels of star formation at large radius, and this process is very stochastic owing to the $\sim 1$ kpc-scale resolution. We also reiterate that our merger definition is quite broad, including mergers completing within $\pm 250$ Myr of the mock image in question and with a mass ratio as high as 10, whereas these diagnostics have been shown to be most useful for mergers within a narrower time window and with mass ratios less than 4 \citep{lotz10}.

While we recover roughly the expected ability of these diagnostics for selecting regions of morphology space containing mergers, the middle column demonstrates that those same regions contain only a minority of the mergers we have defined.  In other words, these selections can be somewhat pure but are very incomplete.  

In the right column of Figure~\ref{fig:classical}, we preview the results of a many-dimensional classification scheme described in the next section, Section~\ref{s:mergers}. While this new scheme includes the regions of morphology space with high merger purity (such as the upper left of the $G$-$M_{20}$ diagram), it also captures the regions of morphology space with lower purity but where the majority of mergers reside.  The distribution of both non-merger and merger morphologies is highly peaked in the center of the diagrams in the middle column of Figure~\ref{fig:classical}, implying that mergers according to our definition do not often look too different from non-mergers when projected in a low number of dimensions.  On the other hand, a many-dimensional classifier appears to be more successful at identifying mergers with subtle features distinguishing them from the non-merger population.

\section{Optimized Image Classifications}   \label{s:mergers}

We seek to derive the best possible image-based classifications for the intrinsic mergers in \illustris. We originally considered two possible approaches: one based on a multidimensional space of summary statistics derived from the synthetic images (manual encoding), and the other based on the synthetic image pixel values themselves (auto-encoding).  An example of the latter is the ``deep-learning'' approach of \citet{Huertas-Company2018} to identify galaxies following evolutionary pathways defined in three-dimensional (3D) physical space. Historically, the former, manual encoding approach uses measurements such as Asymmetry, Gini, $M_{20}$, Concentration, Shape Asymmetry, and others to separate mergers from non mergers. 

While appealing for its potential to robustly disentangle subtle features, the auto-encoding approach suffers from a few possible drawbacks. For one, it is likely that large future imaging surveys such as LSST and Euclid, and large future simulation projects such as the successors to Illustris-TNG, will only analyze images internally and release only catalogs of high level summary statistics. It may not be computationally feasible for external scientists to apply new classifications, such as convolutional neural networks, to large sets of these observations. Therefore, the manual encoding method has the benefit of being accessible to any individual regardless of their infrastructure to efficiently access the underlying pixel data. Another drawback of the auto-encoding approach is that it could require training sets larger than currently available in simulations. Further, there is the danger of training on subtle image artifacts (real or simulated) as opposed to statistics that have been tested extensively on real data. On the other hand, the manual encoding approach is limited to features emergent in the set of summary statistics we choose as inputs. A plus is that we can use algorithms that quantify the performance of the input summary statistics, allowing direct comparison to previous simpler methods used throughout the literature.

In this work, we use a manual encoding method to classify mergers. For classification inputs, we use the results of morphology measurements described in Section~\ref{ss:morphstats}, applied to synthetic images taken in filters common to \hstfull\ surveys of distant galaxies. We have also experimented with measurements on simulated JWST NIRCAM images (highlighted in Figure~\ref{fig:dataset}). We have taken several approaches to filter selection, including using one filter at a time and by combining measurement inputs from two filters, where the specific choices are either fixed over cosmic time or evolved to approximately match the same rest frame wavelengths. When using two fixed filters, we use the filters with most current sky coverage, the HST ACS F814W (I) and WFC3 F160W (H) filters.

\subsection{Random Forests}

To construct summary statistic-based classifiers, we use random forests \citep[RFs;][]{Ho1995}, an ensemble learning method using numerous decision trees to subdivide the space of input summary statistics based on the locations of objects with specified labels in the training set. We use the SciKit-Learn Python code \citep{SciKitLearn2011} to create RFs using the extended algorithm of \citet{Breiman2001}, which uses bootstrapping of the training set and randomized feature selection to control bias, reduce variance, and prevent over-fitting the training set. 

For each simulated image we have one set of morphology measurements from each filter and each camera angle. We discard a small subset of these measurements that fail, for example with very low signal-to-noise per pixel or pathological petrosian radius measurements. We use the remaining sources, numbers listed in Table~\ref{tab:dataset}, for constructing RFs. The input data include the image measurements from either one or two filters per snapshot, where we have experimented with different subsets of morphology parameters. We discuss the choices of input parameters in more detail below, in Section~\ref{ss:features}.  

In addition, we label each source with a binary True or False value according to the intrinsic merger definitions of Section~\ref{ss:mergerdef}. 

{ To construct each random forest, we use 2000 decision trees with a maximum number of leaf nodes selected to achieve reasonable results with a limited degree of overfitting. This maximum number of leaf nodes ranges from $\sim 5$ at $z=4$ to $\sim 50$ at $z=0.5$. For each split in each tree, we use a maximum of four features to determine the best split.}

\subsection{Definitions and Cross Validation}  \label{ss:crossval}

We use cross-validation to tune and verify the parameters of the RF algorithms. For every RF, we select a random subset of $2/3$ of the simulated sources as inputs to construct the forest, and we test the resulting forest classifications against the untrained $1/3$ of sources. We use the test-set results when quoting classification performance.

The output of the RFs is an ensemble of estimates of the probability that a given point in feature space has a given input label.  In our case this output is the probability that the source is a true merger according to Section~\ref{ss:mergerdef}. For each point in the training set and test set, we compute the mean probability from the set of trees in the RF. However, this probability does not refer to a true statistical estimate, but it is instead a variable that could have arbitrary scale depending on the properties of feature space and input data. Therefore, we must choose a method for mapping between these probability values and a classification outcome in order to assess its performance. 

Initially, we explore the space of classification outcomes as a continuous function of the variable values or output probability values. For any chosen threshold value $P$, we assess the performance of the RF classifier using standard definitions. We will also use existing common merger definitions in feature space to make comparable assessments. At a given snapshot, let the number of sources be $N$, the number of mergers be $N_m$, so the number of nonmergers is $N-N_m$. These labels are according to the intrinsic definition of Section~\ref{ss:mergerdef}.  We define the following measures of a classification:
\begin{itemize}
\item{$TP$:  The number of true positives, i.e. the number of true mergers selected by the classification.  $TP \leq N_m$.}
\item{$FP$:  The number of false positives, i.e. the number of non-mergers selected by the classification.  $FP \leq N-N_m$.}
\item{$TN$: The number of true negatives, i.e. the number of non-mergers rejected by the classification.  $TN \leq N-N_m$.}
\item{$FN$: The number of false negatives, i.e. the number of true mergers rejected by the classification.  $FN \leq N_m$.}
\end{itemize}

With these definitions, the total number of objects selected by a classification is the sum of all objects selected, i.e. $TP+FP$.  The total number of objects rejected by a classification is $TN+FN$. The number of true intrinsic mergers is $N_m = TP+FN$, and the number of intrinsic non-mergers is $N-N_m = FP + TN$. These quantities define the ``confusion matrix'' of a binary classification. We then assess the classifications using several common metrics:
\begin{itemize}
\item{ \emph{True Positive Rate} = TPR $= TP/N_m = TP/(TP+FN)$.  This quantity is sometimes called the \emph{Completeness}, \emph{Recall}, or \emph{Sensitivity}. }
\item{ \emph{False Positive Rate} = FPR $= FP/(N-N_m)$.  This quantitiy is also called the \emph{Fall Out} rate.}
\item{ \emph{Positive Predictive Value} = PPV $= TP/(TP + FP)$.  This is the fraction of objects selected correctly, also called \emph{Precision}. A very inclusive classifier can have high TPR but low PPV if it includes a large number of false positives.} 
\end{itemize}

\begin{figure}
\begin{center}
\includegraphics[width=3.5in]{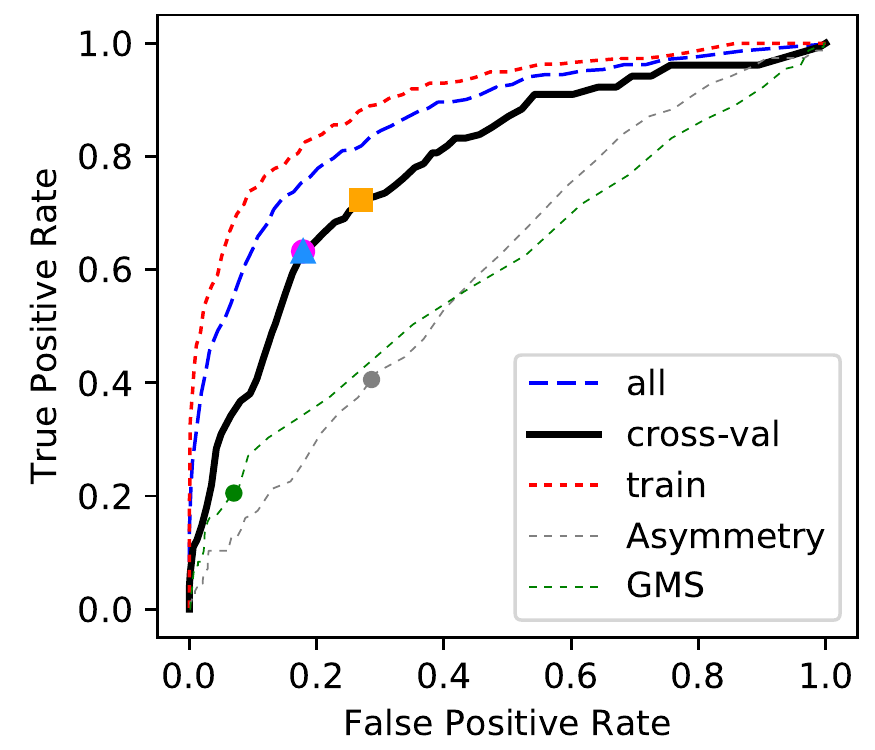}
\caption{Receiver Operating Characteristics (ROC) curve for merger classifications in \illustris\ at $z=1.5$, showing the tradeoff between completeness (true positive rate or TPR), the fraction of true mergers selected, and false positive rate (FPR), the fraction of non-mergers selected. The solid black curve shows the results of the fixed two-filter random forest classifier that we trained and cross-validated against disjoint samples of the image sample (training fraction $0.67$). The green and gray dotted curves show merger samples selected only with Asymmetry or GMS($G$, $M_{20}$) values. \label{fig:roc} }
\end{center}
\end{figure}

Figure~\ref{fig:roc} shows the results of the fixed two-filter (I and H) RF classification for the simulated images at $z=1.5$. Results are similar for simulated sources at different redshifts and different input features, see discussion below for more details. We first use the Receiver Operating Characteristic (ROC) curve, which plots the TPR against the FPR, to assess classification performance. A perfect classifier resides at the upper left of this space, having TPR=1 and FPR=0, and classifications with no discrimination power lie along the diagonal from (0,0) to (1,1). We can then assess classifiers based on the degree to which they approach the upper left region of the ROC plot.

The results of the RF occupy ROC space between the two extremes, indicating that the RF classifications perform well, but not perfectly. For comparison purposes, we plot two classifications using only one or two input features, Asymmetry alone or a diagnostic based on $G$ and $M_{20}$ ($S(G,M_{20})$, Section~\ref{ss:morphstats}). Depending on the thresholds chosen, the RF results have superior TPR or FPR compared to Asymmetry or $S(G,M_{20})$, at the factor of $\sim 2$ level. However, these individual classification schemes were shown to perform best only for a very short period of time surrounding the merger event, and therefore it is unsurprising that according to this metric, they perform less well for this more difficult target set of minor and major mergers occurring within a window of 500 Myr around the observation time.

\begin{figure}
\begin{center}
\includegraphics[width=3.5in]{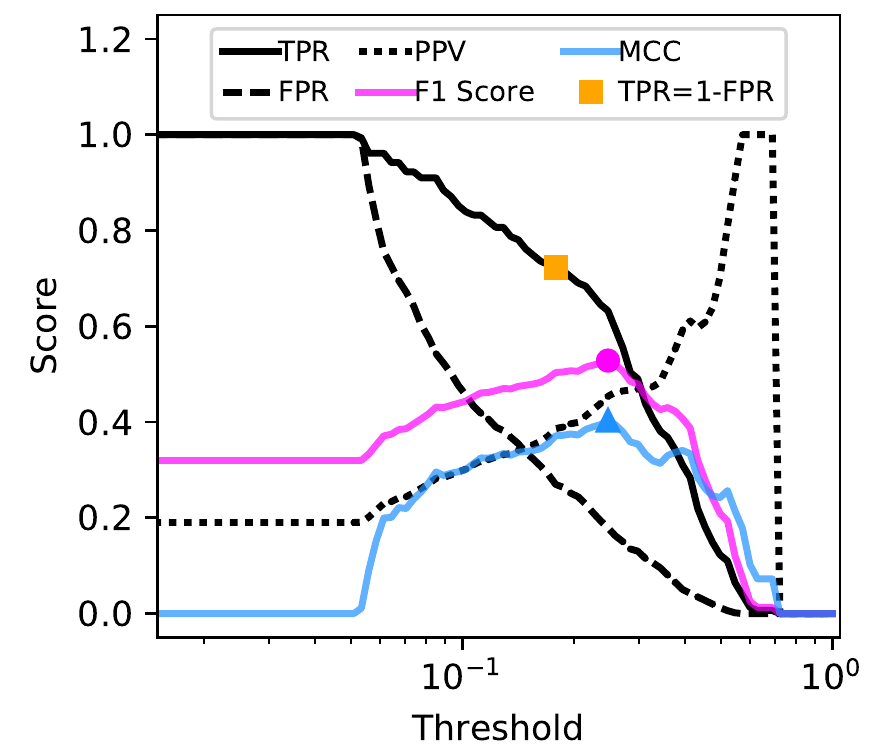}
\caption{ Binary classification metrics for the two-filter classifier as a function of the probability assigned to each object by the random forest tree. The black dashed curve shows the sample purity (positive predictive value or PPV), the fraction of selected objects which are true mergers. The blue, orange, and purple curves show three alternative methods for selecting an optimal threshold for defining the classifier, the balance point ($TPR=1-FPR$, orange), the inverse F1 score (magenta), and the Matthews Correlation Coefficient \citep[MCC, blue;][]{Matthews1975}. \label{fig:rocthresh} }
\end{center}
\end{figure}

\subsection{Selecting a Probability Threshold}  \label{ss:probthresh}

\begin{figure}
\begin{center}
\includegraphics[width=3.5in]{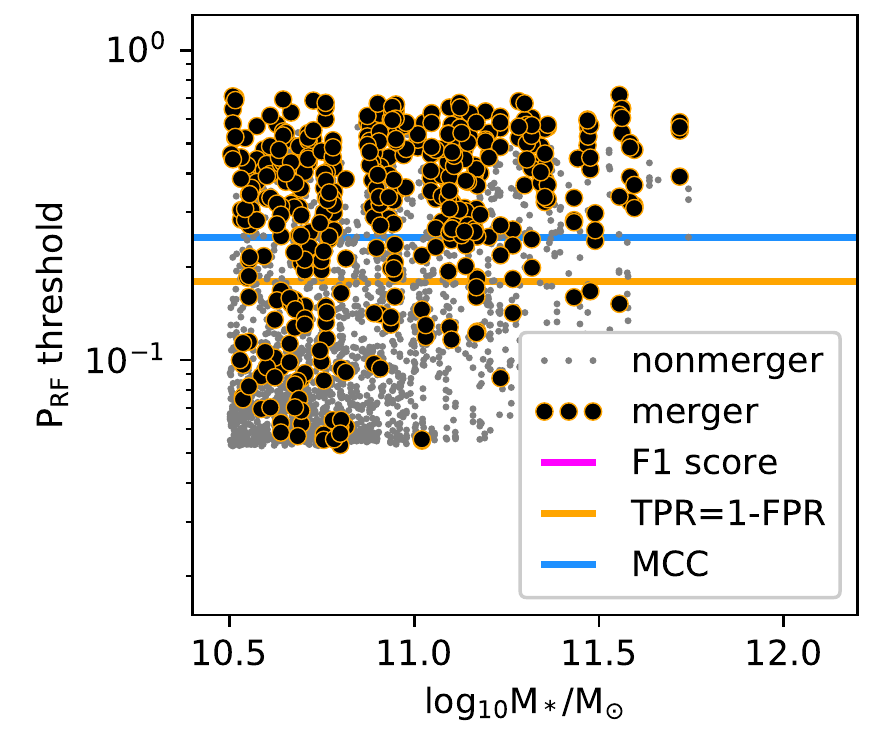} \\
\includegraphics[width=3.5in]{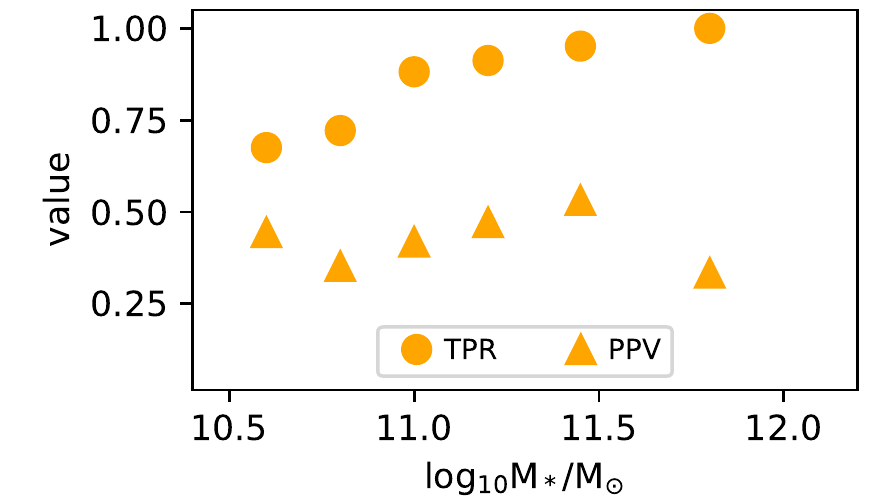}
\caption{Top: Merger probability assigned to each \illustris\ source at $z=1.5$ by the fixed two-filter RF classifier, as a function of $M_*$. This presents the same information as Figures~\ref{fig:roc} and \ref{fig:rocthresh} but in a slightly different form. Solid horizontal lines show the probability thresholds chosen by optimizing the three functions shown in Figure~\ref{fig:rocthresh} (The magenta line is covered by the blue line). In this paper, we use the orange line to define the merger classification threshold. This is the probability value for which $TPR=1-FPR$, in other words the fraction of true mergers selected equals the fraction of non-mergers rejected (completeness equals specificity). {  Bottom: Completeness (TPR) and purity (PPV) as a function of stellar mass for the RF results from the top panel.} \label{fig:probmass}}
\end{center}
\end{figure}

The ROC curve is a common way to evaluate the performance of binary classifications as a function of the threshold value, which can be adjusted to achieve a chosen performance along the curve \citep{Fawcett2006,Powers2011}.  However, by itself this analysis does not demand that any single location along the curve is obviously the best choice for a given problem. Therefore, we explored several methods for selecting a single optimum point along the ROC curve as the one classification scheme for a given RF. This approach provides flexibility in selecting the best threshold value, which we choose to let vary as a function of redshift and as a function of the input features and RF tuning parameters.  This adds complexity to our accounting of the classification schemes, because we must keep track of the results of the threshold-optimizing algorithm, but it is valuable for comparing the RF and classical classifications with each other, across time, and with varying assumptions. 

The three summary statistics we explored for the RF classifications include:
\begin{enumerate}
\item{The F1 Score: the harmonic mean of the TPR and PPV: $ F1= 2 \frac{TPR \times PPV}{TPR + PPV}  $.  F1 occupies 0 (worst) to 1 (best). }
\item{The Balance Point: the point at which TPR=1-FPR. This is the point where the ROC curves intercept a line drawn from (0,1) to (1,0), which is orthogonal to the no-discrimination line. }
\item{Matthew's Correlation Coefficient \citep[MCC][]{Matthews1975}: MCC $= (TP \times FN - FP \times FN)/\sqrt{(TP+FP)(TP+FN)(TN+FP)(TN+FN)}$.  Widely used in bioinformatics, some regard MCC as an excellent measure of the quality of a binary classification, performing well even when the two classes have very different sizes \citep{Boughorbel2017}, an important property when mergers are rare. A perfect prediction has $MCC=+1$, a classifier performing no better than random has $MCC=0$, and a prediction that perfectly disagrees with data has $MCC = -1$.}
\end{enumerate}

In Figure~\ref{fig:rocthresh}, we show how these scores, as well as TPR, FPR, and PPV, depend on the RF output probability threshold for the example snapshot at $z=1.5$ with two fixed filters (I and H). Recall that the balance point occurs for TPR=1-FPR. For the F1 and MCC scores, we select the RF output probability threshold that maximizes the score, and we plot the three optimum values as the solid points with corresponding colors and shapes in Figures~\ref{fig:roc} and \ref{fig:rocthresh}. We inspected these plots for each of the redshifts and filter choices considered and found that often, the three methods select probability thresholds that are similar to each other (within roughly a factor of $2$). Figure~\ref{fig:probmass} shows how these probabilities and thresholds depend on stellar mass.

From this point forward, we will quote results using the Balance Point to select the final probability threshold and thus define the classification based on each RF. The balance point often occurs for lower threshold values, and therefore this method leads to higher sensitivity (TPR) at the expense of slightly weaker precision (lower PPV). Therefore, RFs at the balance point tend to select a greater number of sources as mergers than the same RF maximizing F1 or MCC.

For comparison purposes, we fix $A \geq 0.25$ and $S(G,M_{20}) \geq 0.10 $ as the threshold levels for these two diagnostics and plot those values on the ROC curves of Figure~\ref{fig:roc}. From this Figure, we see that if we chose instead to let the threshold values vary in the same way as the RFs, choosing the balance point summary statistic (presumably also F1 or MCC), the values of the TPR for A and S will increase to be $\sim 0.5$, still somewhat less than the RF at $\sim 0.7$. However, with this assumption, their FPRs increase substantially to $\gtrsim 0.4$, higher than the RF at $\sim 0.3$-- in other words, they become noticeably less pure than the RFs.  As we show in Section~\ref{s:mergerrates}, this general property holds for other filter choices and redshifts.

\subsection{Discussion of Feature Importances}  \label{ss:features}

\begin{figure}
\begin{center}
\includegraphics[width=3.3in]{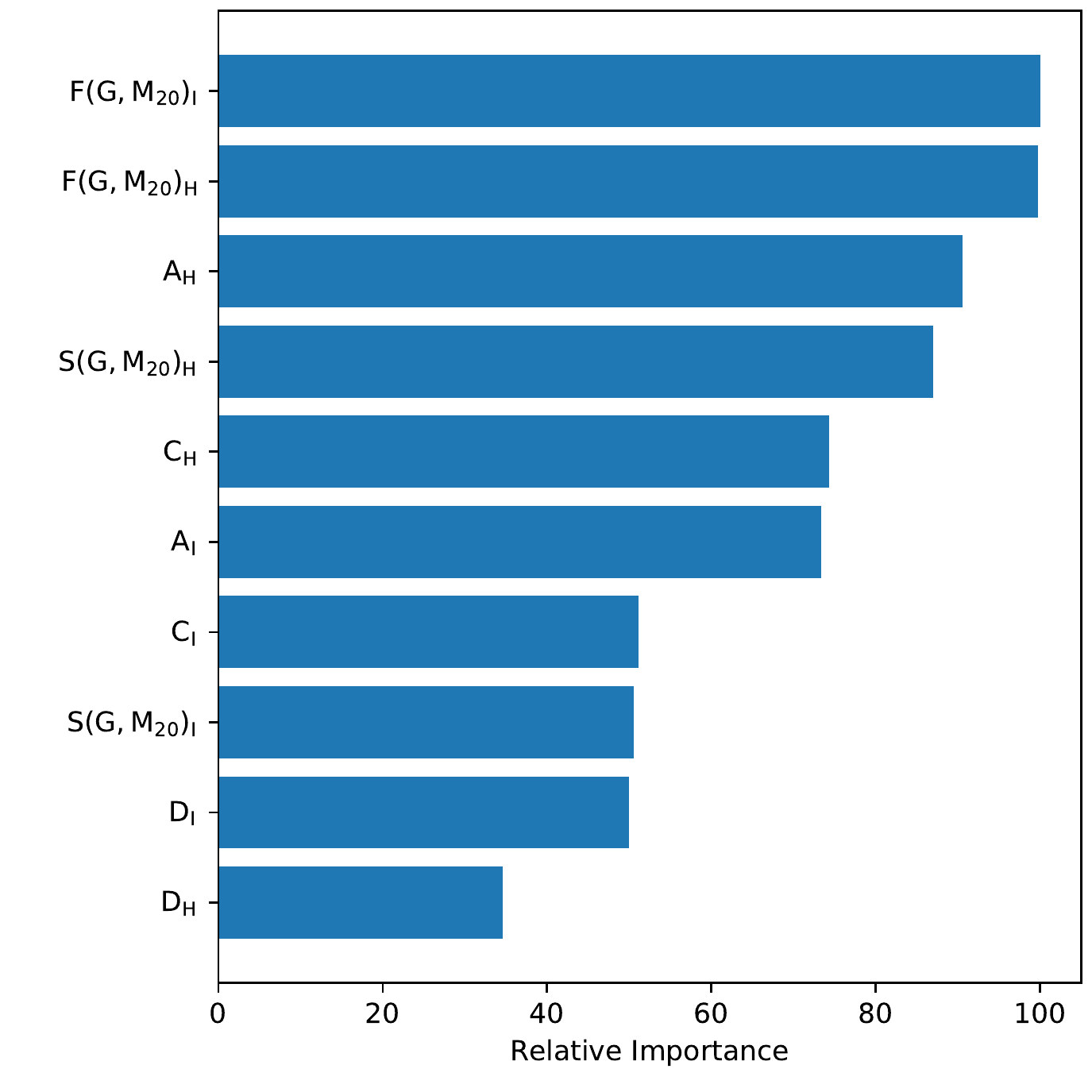} \\
\caption{Relative feature importances for the fixed two-filter RF classifier defined at $z=1.5$. This demonstrates that the merger selection depends strongly on the total underlying morphology: measures correlating with bulge strength ($F(G,M_{20})$, C) and disturbances (A) both have high importance. \label{fig:features} }
\end{center}
\end{figure}

So far, we have shown RF results using a particular set of input features measured from images in two filters (I and H), which will be our primary definition of RF classifications going forward. We came to this decision through trial-and-error, by creating RFs with various sets of input parameters measured from one or two filters chosen in different ways, either fixed observed filters or evolving to roughly match the same rest-frame wavelengths. For each trial, we visually inspected the ROC curves and threshold plots (corresponding to Figures~\ref{fig:roc} and \ref{fig:rocthresh}) in order to understand how these variations changed the quality of the resulting classifications.

Some trials included adding input parameters to the RFs, such as information about the intrinsic state of each galaxy, for example stellar mass, halo mass, or SSFR. These trials tended to improve the classification performance, although for sometimes pathological reasons. For instance, because we measure sources from multiple camera angles that all have the same intrinsic property to machine precision, the RF sometimes learned which exact values corresponded to mergers based on having images of the same galaxy split into both the training and test set. In addition, in \illustris\ at least the merger incidence is a strong function of mass, and so including these parameters can lead to good classifications without indicating beneficial discriminatory power from the image morphologies.  For these reasons we do not use such classifications further.

Figure~\ref{fig:features} presents our default set of input features and their relative importance to the RF classification outcomes, defined as the mean decrease in impurity achieved by each variable at all relevant nodes in the forest \citep[for more, see e.g.,][]{Breiman2001,Freeman2013}.  We find that the classification depends on the total morphological signature, including not only long-known important measures of disturbances (Asymmetry, $S(G,M_{20})$, $D$) but also the extent to which the source is dominated or not by a central bulge. We discuss this interesting phenomenon further in Sections~\ref{ss:pastfuture} and \ref{s:discussion}.  As shown in Figure~\ref{fig:classical}, this enables the RF classifier to be sensitive to mergers throughout the space, and not just the ones that are blue or disc-dominated as common indicators select.

Other trials swapped different image measurements, for example we found that including the $M$ and $I$ statistics of \citep{Freeman2013} did not greatly improve the classification performance. Furthermore, these statistics had the lowest relative feature importances ($\lesssim 20\%$) when included in plots such as Figure~\ref{fig:features}. This result might owe to slight differences in the ways we measured these statistics compared to their original formulation, including differences in image segmentation algorithms, and so we defer to later a more robust investigation of these statistics. 

Another choice involved keeping or removing the average signal-to-noise ratio (SN$_{\rm pix}$) from the RF inputs, which we hypothesized might be useful for tracking performance as a function of image quality. This parameter had low but non-trivial importance to the classifications, and we found that removing it did not worsen the ROC curves or statistical scores of the RFs in a noticeable way. We choose to not include these parameters in our default formulation of the RF classifications, to avoid classifying on extra information about, for example, mass or colour.

We investigated many options for choosing from which images to base the RF inputs, drawing from the set of different instruments and filters common to HST (and planned JWST) surveys we used to create the mock images. Our default choice is to use I and H band images, common to all CANDELS fields, for the RF input measurements. In general, switching the redder filter to NIRCAM from WFC3 did not perceptibly enhance the RF classifications, nor does using the evolving filter set highlighted in Figure~\ref{fig:dataset} (see Figure~\ref{fig:rocevolve}).

\begin{figure*}
\begin{center}
\includegraphics[width=6.5in]{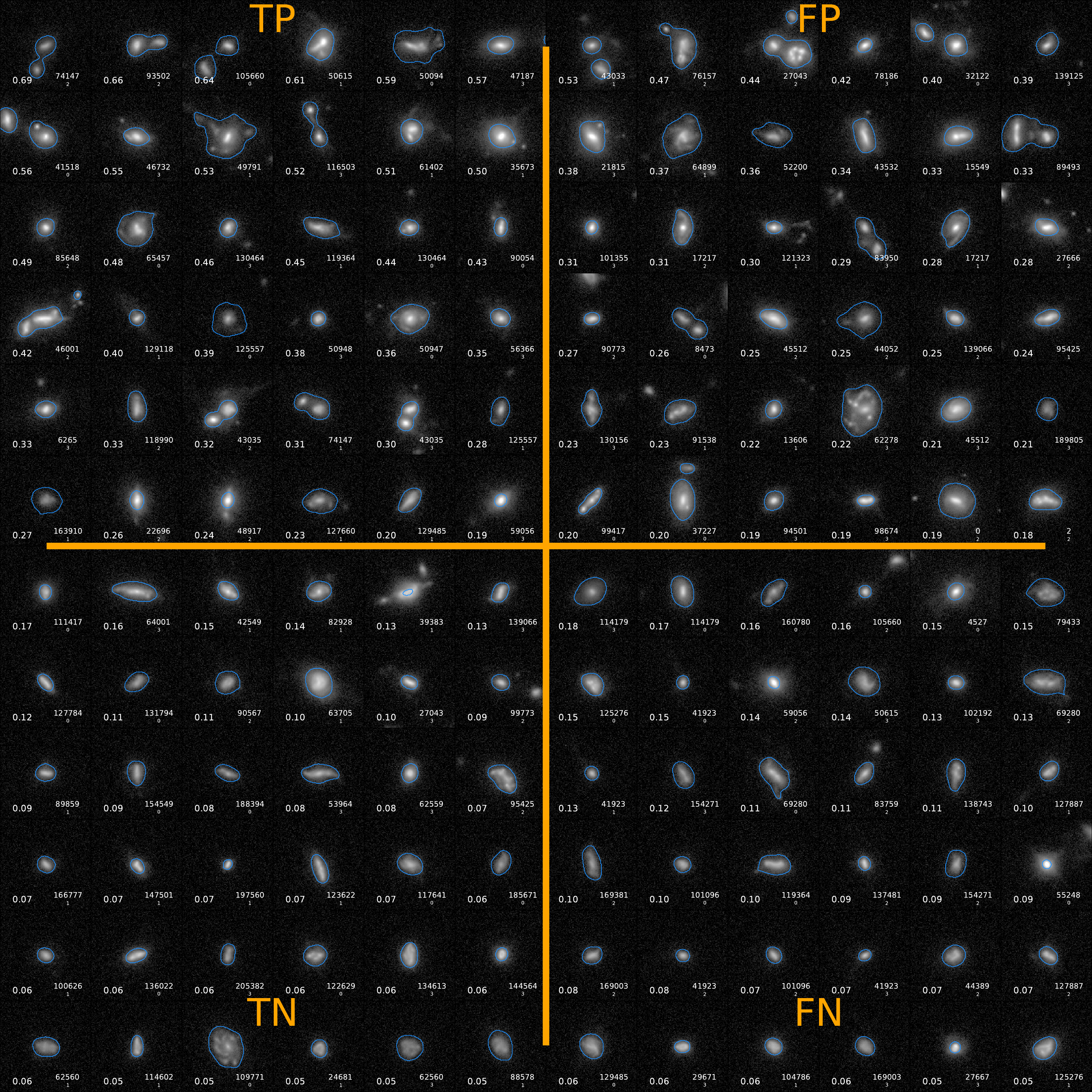}
\caption{Grid of F160W images organized by selection result for the fixed two-filter RF classifier defined at $z=1.5$. The upper left shows true positives (TP, true mergers selected), upper right false positives (FP, non-mergers selected), lower left true negatives (TN, non-mergers rejected), and lower right false negatives (FN, true mergers rejected). Thus the horizontal boundary separating the top and bottom halves of the figure corresponds to the RF probability threshold selected in Section~\ref{ss:probthresh}. Each of the four quadrants shows sources sorted and sampled evenly by the output RF merger probability.  Each source image lists the RF probability, the \illustris\ subhalo ID number (snapshot 075), and the camera number.  The blue outlines in each panel show the segmentation map generated during the morphology measurement process \citep[following][]{Lotz2004}, in order to judge when blending might contribute to the classification results.  \label{fig:imageclass} }
\end{center}
\end{figure*}

\subsection{Classification Result Examples}

In Figure~\ref{fig:imageclass} we provide a summary of simulated sources that have been classified by our default RF.  In each quadrant, we show 36 H-band images sampled evenly among the output RF probabilities in each of the four categories TP, FP, TN, FN.   We supply corresponding figures for each of the snapshots considered in this work as online-only supplementary material.  

In addition, we used these figures as a tool to visually diagnose and debug the RF classification schemes, beyond the assessments provided by the ROC analysis.  For each of the sources in Figure~\ref{fig:imageclass}, we created a plot like those shown in Figure~\ref{fig:merger_examples}: a zoomed image alongside the stellar mass growth history, and highlighting the next and prior major and minor mergers, respectively. Using specific image formats (.svg) and hyperlinks embedded via Matplotlib \citep{Hunter:2007}, we created a web-friendly set of files, where a user can click each panel of the image grids to evaluate the correctness of the classifications and explore why certain sources were classified as they were. We include each of these individual plots as supplementary material. 

\begin{figure}
\begin{center}
\includegraphics[width=3.3in]{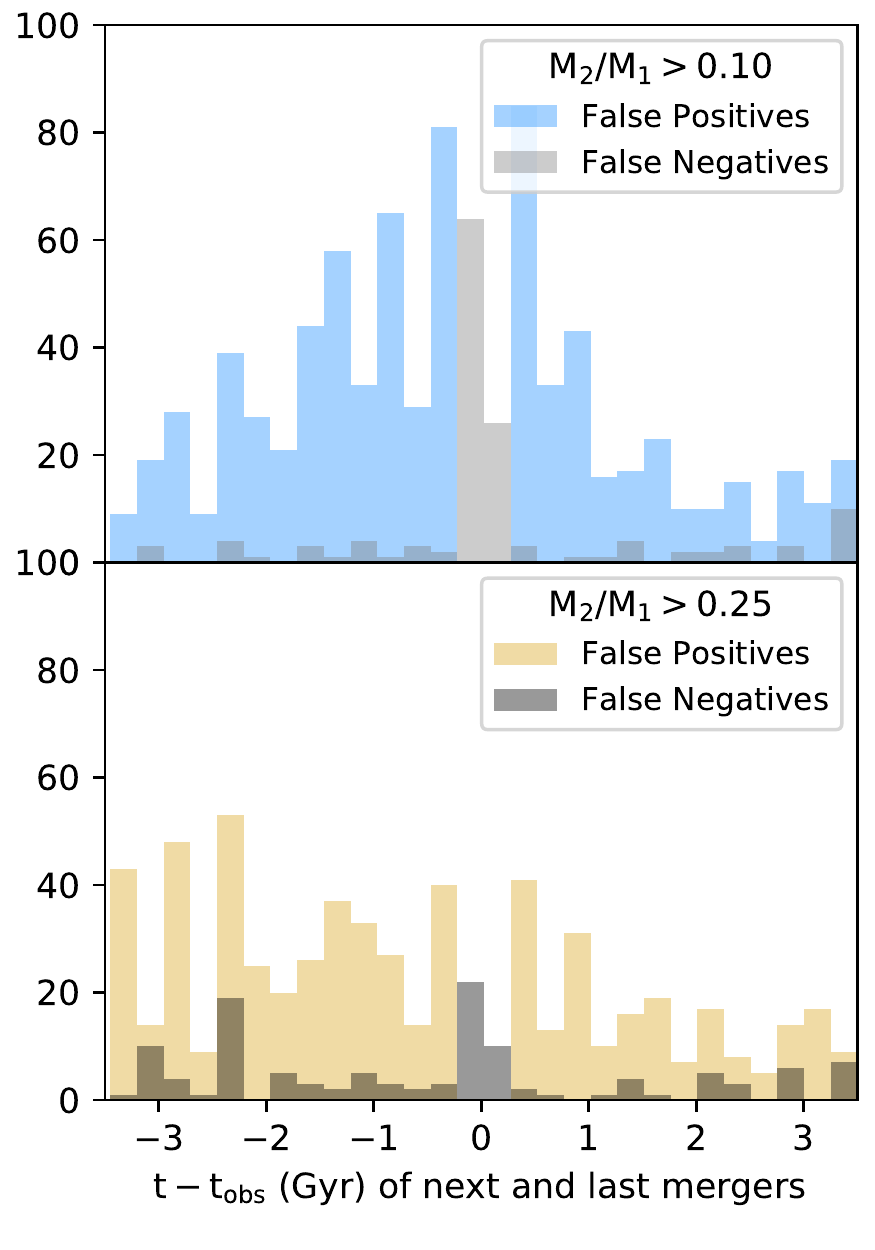}
\caption{Distribution of merger event times relative to each mock observation, for each source classified as a False Positive or False Negative in our default RF for $z=1.5$. Top: Histogram of the time of next and last major or minor mergers considered in the RF.  Bottom:  For the same sources, the time distribution of the next and last major mergers only. False positives have a large contribution from both major and minor mergers occuring within a few Gyr of the observation (but outside the 0.5 Gyr window), as well as seemingly non-merger-related processes such as clumpy star formation. False negatives are dominated by minor mergers.  \label{fig:class_stats} }
\end{center}
\end{figure}

From these images, we can investigate further how well the RF classifications perform, especially to explore their failure modes. We find that many False Positives are triggered when a merger event occurs outside (but possibly near) the 500 Myr time window we used to label mergers in Section~\ref{ss:mergerdef}. Therefore, many such failures are caused by projections or interactions not leading to an immediate merger. This type of failure will occur in any classification with such a narrow definition of merger time (at coalescence), but it may be possible to capture these sources in a broader classification accounting for the fact that they might merge on a wide range of times relative to the observation time.  The False Negative failure mode results largely from minor mergers failing to trigger long-lived or perceptible morphological disturbances.  In Figure~\ref{fig:class_stats}, we quantified the distribution of merger events related to these two failure modes, verifying the conclusions above.

\begin{figure}
\begin{center}
\includegraphics[width=3.3in]{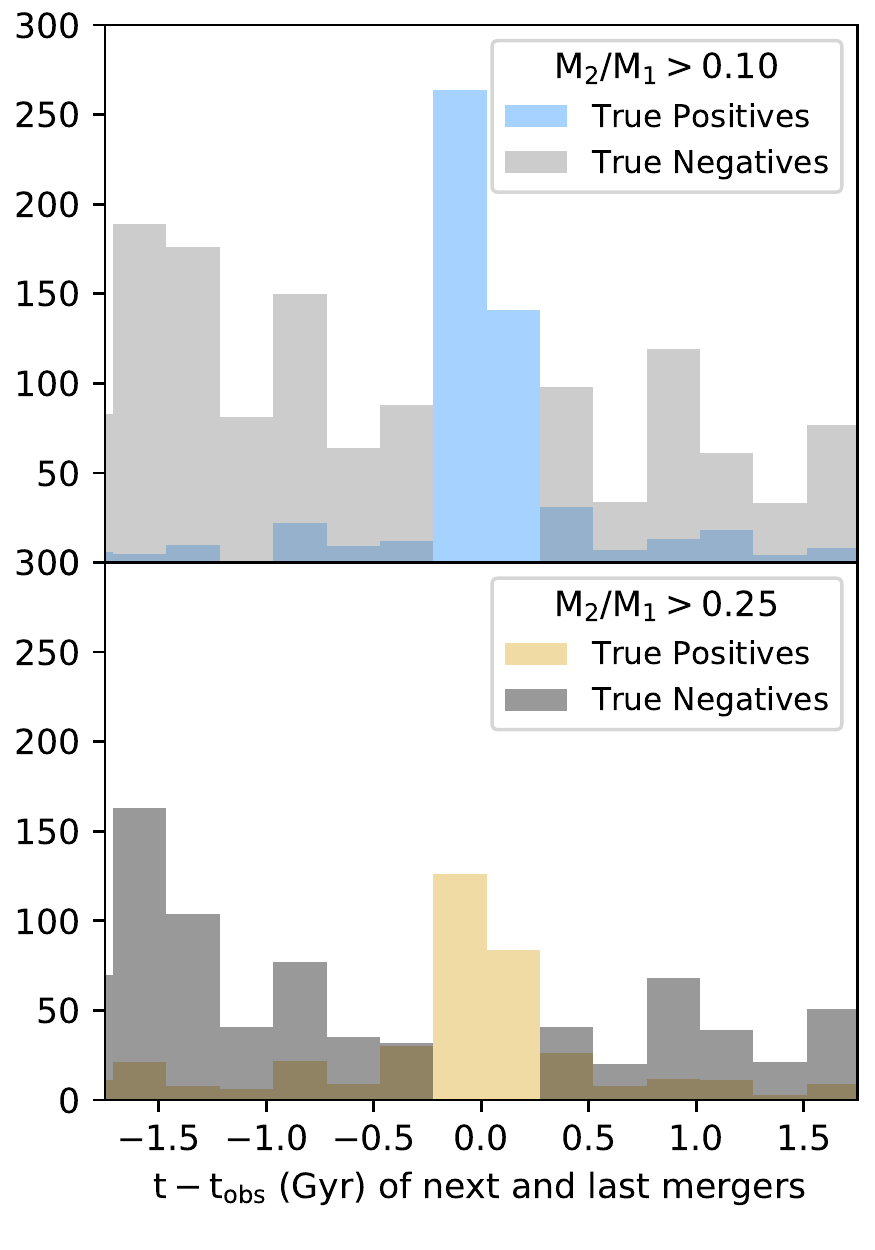}
\caption{   Same as Figure~\ref{fig:class_stats} but for True Positive and True Negative sets.   \label{fig:class_true_stats} }
\end{center}
\end{figure}

Figure~\ref{fig:class_true_stats} shows the same information for the True Positive and True Negative classifications from the same set. We plot bin sizes of $250$ Myr to capture the relative contribution of past and future mergers. From these, we find that the RF selection is capturing a significant number of mergers completing both before and after the time observed, with $\sim 75\% $ more True Positive mergers completing in the past than the future.

\subsection{Features Selecting Past vs. Future Mergers} \label{ss:pastfuture}

We tested which features were responsible for selecting past (future) mergers by defining two new training sets composed of just mergers completing within the past (next) $250$ Myr of observation time. We re-created RF classifications with the same settings as our default RFs and use ROC analysis to find that both sets achieve some success in training viable classifications, albeit less success than the full $500$ Myr-wide training set. We conclude that the combined RFs are able to select a wide range of merger stages.  

\begin{figure}
\begin{center}
\includegraphics[width=3.3in]{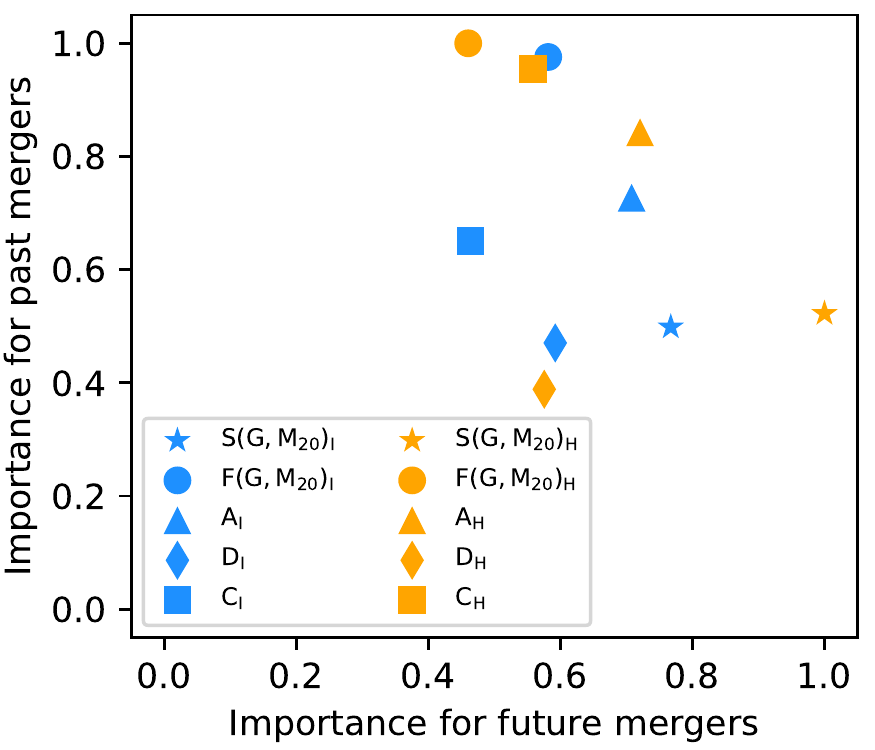}
\caption{Relative feature importances for RF classifications trained on sets of mergers completing within the past $250$ Myr plotted versus mergers completing within the next $250$ Myr, i.e. the past and future subsets of the full default RF training sets used in the rest of this paper. Signatures of a bulge contribute to selecting recent mergers, while signatures of asymmetry and multiple nuclei contribute to selecting mergers completing soon. \label{fig:pastfuture}}
\end{center}
\end{figure}

In Figure~\ref{fig:pastfuture} we highlight which morphology features were important for identifying past-only or future-only mergers in the split training sets.  We find that metrics commonly associated with a strong bulge component ($C$, $F(G,M_{20})$) contribute most to selecting a pure set of past mergers.  Metrics associated with disturbances ($S(G,M_{20})$ and to a lesser extent $D$) contribute to selecting future mergers. Asymmetry $A$ is important for both classes.

\section{Evolution of Merger Statistics}  \label{s:mergerrates}

In this section, we summarize the results of the random forest (RF) classification method applied to each snapshot from the entire mock image dataset. Of the eight \illustris\ snapshots we considered at $0.5 \leq z \leq 5$ using our default noise parameters, only the seven at $0.5 \leq z \leq 4$ yielded enough detections to construct a training set for RF classifications. For each RF we construct, we follow the procedures described in Section~\ref{ss:probthresh} to fix the optimal probability threshold at the balance point. We store this value along with the RF decision trees for each snapshot and choice of filters. We can then use these objects to apply the RF classifications so defined to any new input set of data with the same input morphology statistics, including other simulated data or real data.

The ideal many-dimensional selection boundaries defined by the RF evolve somewhat as a function of redshift. We investigated the idea of combining the entire set of \illustris\ mock images into a single set of images spanning the whole redshift range. This would allow one to define a single RF classification for galaxies at all epochs, and could yield benefits from the increased size of the training set.  However, we found that the individually trained RFs, which can vary arbitrarily with redshift, performed better than the combined RF. We hypothesize that this result owes to strong evolution in important parameters of the merger identification process, such as angular size evolution, noise, morphological change, or evolution of the merger rate. Thus it is important to define flexible classification schemes that can evolve at least as fast as these observational and physical effects.

\subsection{Redshift Evolution of Classification Performance}

\begin{figure*}
\begin{center}
\includegraphics[width=3in]{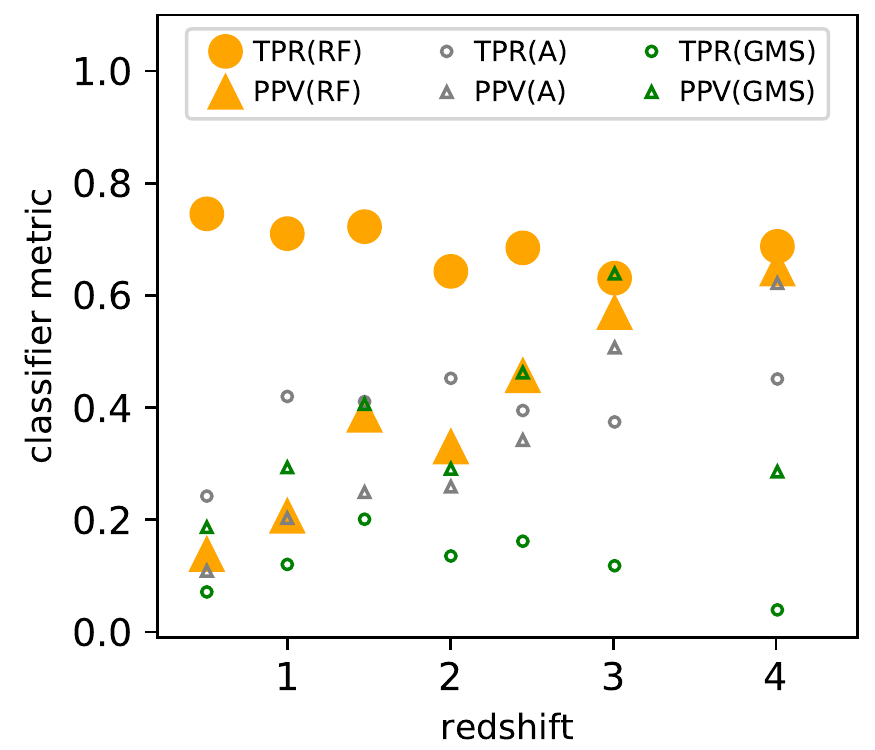} \includegraphics[width=3in]{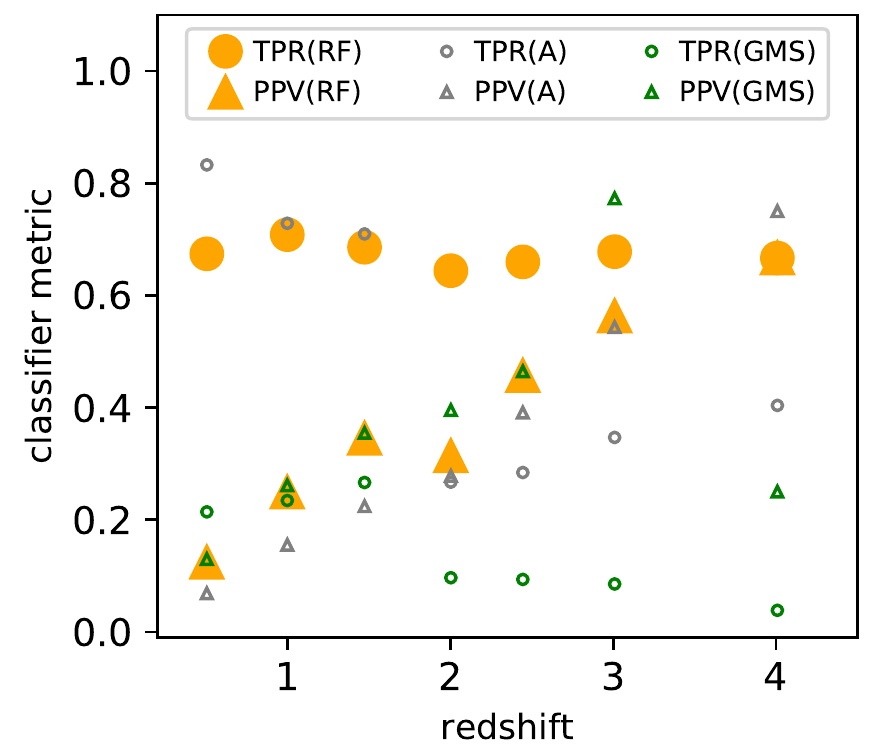}
\caption{Redshift evolution of RF classification results for the fixed (left) and evolving (right) two-filter classifers. At all times, the RF classfier selects a highly complete ($TPR \sim 70\%$) merger sample, with a purity that improves with distance to $PPV \sim 50\%$ at $z=3$. We also show results for one-filter selections on Asymmetry and $GMS(G,M_{20})$. Generally speaking, the two-filter RF classifier exhibits much greater completeness TPR and modestly better (A) or similar (GMS) purity PPV.  For any selection technique, these values are necessary to compute the intrinsic merger fraction given a classified sample. \label{fig:rocevolve} }
\end{center}
\end{figure*}

In Figure \ref{fig:rocevolve}, we show how the RF performance changes as a function of redshift, using the metrics defined in Section~\ref{ss:crossval}.  Here we show results from both the fixed two-filter RF scheme discussed in Section~\ref{ss:features} (I and H only), as well as using the evolving pair of filters shown in Figure~\ref{fig:dataset}. We find consistently good completeness, $TPR \approx 0.7$, and a purity that increases from $PPV \approx 0.2$ at $z=0.5$ to $PPV \approx 0.5$ at $z=3$. 

For comparison, we plot the summary of the fiducial Asymmetry-only and $G$-$M_{20}$-only classifiers (Section~\ref{ss:probthresh}) applied to the bluer filter of each pair.  These statistics have been shown to have shorter observability time ($\lesssim 200$ Myr) than the time window defining our merger training set ($500$ Myr), and their sensitivity to mergers of different mass ratios is variable. { Indeed, we find that the completeness of the Asymmetry and $G$-$M_{20}$ statistics improves by roughly a factor of $1.5$-$2.0$ when using a 250 Myr instead of 500 Myr time window. }  Therefore, Figure~\ref{fig:rocevolve} only demonstrates the relatively better completeness of the RF on this broader, more challenging training set, and do not imply that these statistics cannot be used to classify mergers defined in different ways. When considering the dataset as a whole, we find either choice of filters leads to similar RF performance, while the individual statistics depend strongly on the filter choice: Asymmetry alone is highly complete ($\approx 0.7$) when measured on the ACS B and V filters (F435W and F606W) for simulated sources at $z < 2$ but much less complete when measured in redder filters (such as I or J). 

We find similar or better values and evolution in the classification purity (PPV) attained by the RF classifier compared to the individual statistics. Given its high TPRs, we conclude that the RFs offer superior ability to identify mergers across $0.5 \leq z \leq 4$.

\subsection{Importance of Total Morphology}

\begin{figure}
\begin{center}
\includegraphics[width=3.5in]{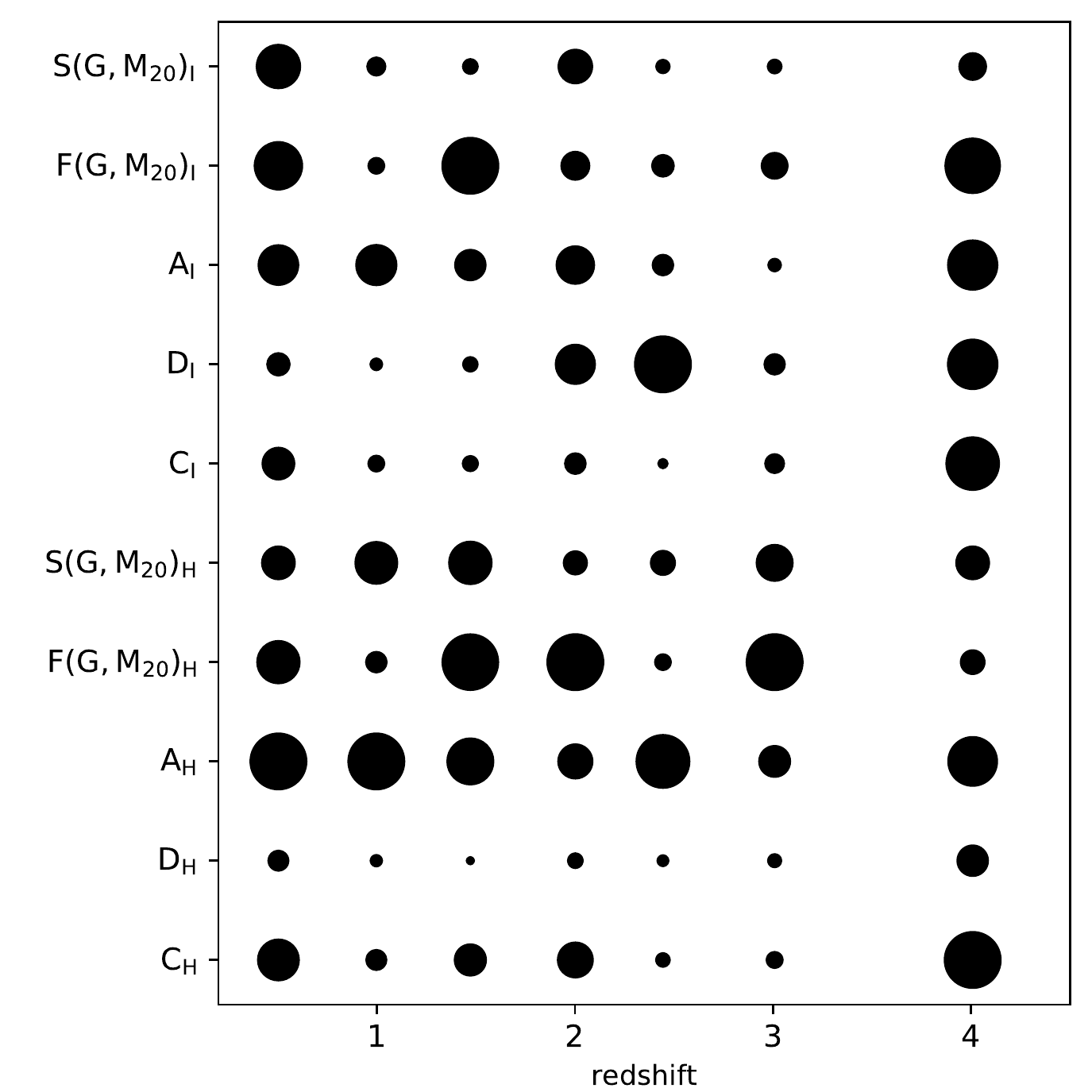}
\caption{Redshift evolution of relative feature importances for the fixed two-filter RF classifiers.  The diameter of each circle is proportional to the relative importance of each feature in each of the independent classifiers. Thus the area is proportional to the square of feature importance. The importances are somewhat consistent from snapshot to snapshot, indicating that it is important to include a wide range of different diagnostics to identify distant mergers. \label{fig:totalimportance} }
\end{center}
\end{figure}

Figure~\ref{fig:totalimportance} shows that the relative feature importances change somewhat as a function of redshift for our fixed two-filter RFs, but there is clear consistency in the feature importances. We see that $F(G,M_{20})$, generally used to measure bulge strength, and Asymmetry from both filters tend to be important in determining the RF classifications at a handful of the snapshots.  H-band $S(G,M_{20})$, generally used to identify mergers with multiple nuclei, and $C$ are also somewhat important at a range of time-steps.  These results are similar to the ones found at $z=1.5$ only (as in Figure~\ref{fig:features}).

This strengthens the conclusions from Section~\ref{ss:features} that it is necessary to incorporate as much morphological information as possible when selecting mergers. Another consequence is that high RF probabilities map out complex manifolds in standard 2D diagnostic diagrams, more closely matching the true distribution of mergers (see Figure~\ref{fig:classical}). Therefore even when analyzing sources at a single redshift, it is inadvisable to use any fixed, low-dimensional classification boundary.  

\begin{figure*}
\begin{center}
\includegraphics[width=3in]{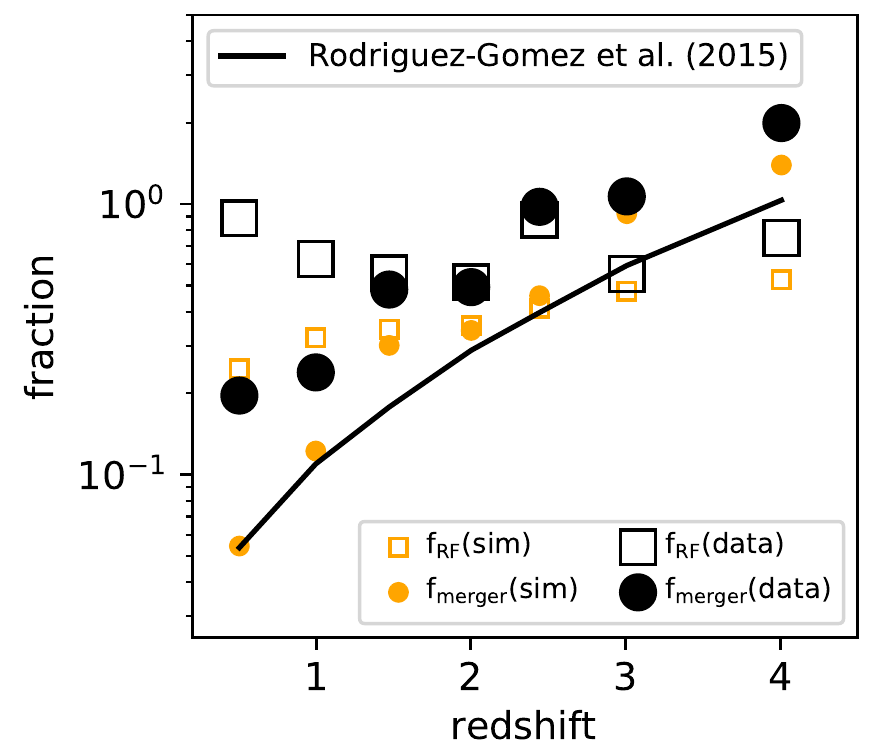} \includegraphics[width=3in]{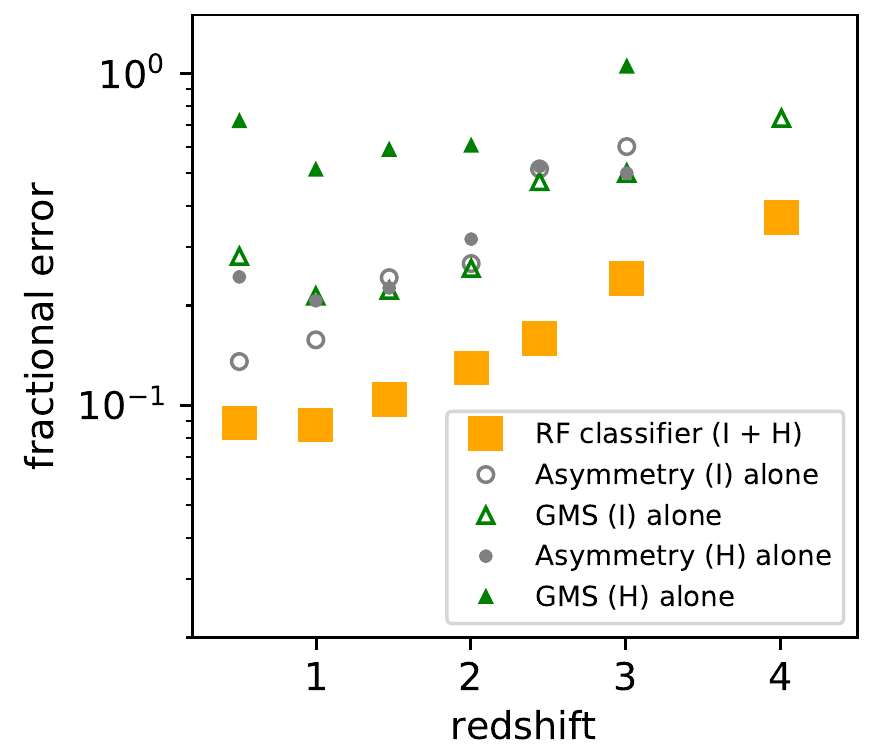}
\caption{HST-based merger fraction (left) and statistical error in the merger fraction (right) versus redshift, applying the \illustris-derived fixed two-filter RF classifiers. At each time, this shows the estimated number of mergers with mass ratio greater than 10:1 within the past or next $0.25$ Gyr relative to the total number of galaxies with $M_* \geq 10^{10.5} M_{\odot}$. The solid curve shows the merger fractions predicted by R-G15 for such galaxies. \label{fig:fractions}}
\end{center}
\end{figure*}

\subsection{Merger Incidence from CANDELS Images} 

Finally, we use the trained RF classifiers to develop observational diagnostics of the merger fraction and merger rate in distant galaxies. For input data, we use morphology catalogs measured by \citet{Peth2016} and Lotz et al. (2018, in prep.) from all five CANDELS fields. Using the multi-wavelength catalogs \citep[e.g.,][]{Galametz2013}, we select sources with SED-based fitted $M_{*} > 10^{10.5} M_{\odot}$ and use photometric redshifts to divide the data into bins surrounding each of the \illustris\ snapshots we considered. We use I and H band measurements of the same morphology diagnostics used to train the \illustris-based RFs, and place the same requirements on measurement quality: no flags and average signal-to-noise per pixel $> 3$. The depths of the training images and CANDELS are similar, but not identical. We used a new Python code for measuring the morphology diagnostics, but we showed that the results are identical to the original IDL code for these studies \citep[e.g.,][]{Lotz2004,Peth2016}.

Figure~\ref{fig:fractions} presents the merger fraction derived from the fixed two-filter RFs in \illustris\ and in CANDELS. We begin with the number of galaxies selected by the RF trees, $N_{RF}$, using the adopted probability threshold. We multiply the selected fraction, $f_{RF}= N_{RF}/N$, by $PPV/TPR$ using the simulation cross-validation procedure described in Section~\ref{ss:crossval}.  This applies the known incompleteness and purity of the derived classifier to correct the raw statistics and obtain the best-guess merger fraction for newly input data. We then multiply the result by $\left < M/N\right >$, the average total number of simulated mergers divided by the number of galaxies with at least one such merger. This extra predicted value is important to achieve self-consistency when counting mergers, because a binary classification cannot distinguish between one and more than one merger event, whereas theoretical predictions usually count the total number of mergers. { For completeness, the full expression for the merger fraction estimated using RF-based selection is:
\begin{equation} \label{eq:mergerfrac}
f_{\rm merger} = \frac{N_{RF}}{N} \frac{PPV}{TPR} \left < M/N \right >.
\end{equation}
The PPV/TPR term corrects for flaws in the classifier using its known training results. Therefore this expression depends sensitively on both the definition of the training set, as well as the quality of the resulting classifications at each epoch. We explore this sensitivity in Section~\ref{ss:newtraining}.} 

We computed the theoretical fraction by multiplying the \illustris\ merger rates (R-G15) by $0.5$ Gyr, the time window for considering an image to be a true intrinsic merger defined in Section~\ref{ss:mergerdef}. We also confirmed this is a reasonable value by inspecting the distribution of $t-t_{\rm obs}$ for simulated mergers selected by the RF and found that it is consistent with being nearly uniformly distributed over the $500$ Myr window. For simulated data, we achieve close consistency between the estimated merger fraction (filled orange circles) and the intrinsic merger fraction (solid black curve) in Figure~\ref{fig:fractions}.  While this occurred by construction, it is an important check of our arithmetic and assumptions.  

It is possible that the image training set could double-count some mergers, so it may be necessary to correct merger fractions estimated from the RFs by a factor smaller than (but close to) 1. The fact that the estimated simulation merger fraction is slightly greater than the intrinsic merger fraction in Figure~\ref{fig:fractions} could result from this effect.  For mergers completing after the image time-step, the \illustris\ image dataset can in principle include an entry for both progenitors.  Therefore, when adding up all mergers, for some sources this effect could add two mergers where the intrinsic merger rates only count one.  This will only happen for future mergers (ranging from $\sim 1/2$ to $\sim 1/3$ of those selected, see Figure~\ref{fig:class_true_stats}) and only when the primary is massive enough that the secondary also enters the RF sample. Therefore, while we are not certain what correction factor to apply, it is at least $\sim 3/4$ and probably $\sim 5/6$, similar to the slight offset in the orange circles and black solid line in Figure~\ref{fig:fractions}.  When applying the RFs to CANDELS data, this effect is likely to affect even fewer sources than for the simulation estimates, because the CANDELS catalogs generally will not double-count very blended late-stage mergers.

Using the RF classifiers, we find the measured merger fraction follows a very similar trend to the predicted merger fraction, rising steeply and continuously from $f_{\rm merger} \sim 10\%$ at $z=1$ to $f_{\rm merger} \sim 100\%$ at $z=4$. However, it is elevated by a factor of $\approx 2$ compared with theory. We discuss possible causes for this difference in Section~\ref{s:discussion}. The plot on the right of Figure~\ref{fig:fractions} shows that because it achieves high completeness (TPR), the multidimensional RF classifier has superior statistical performance in estimating merger fractions, with fractional errors $\sigma_{f}/f \approx 10\%$ at $z=2$, about a factor of 2 smaller than classical one-filter diagnostics.

{

\subsection{Choice of Training Set} \label{ss:newtraining}

 Throughout this paper, we used a training set with a distribution of galaxy morphologies and merger incidence reflecting the intrinsic \illustris\ galaxy population.  Given that non-mergers typically outnumber mergers by a fair margin in these samples, we have therefore used a highly unbalanced training set to create the RFs. While we have used RF techniques that attempt to mitigate serious biases caused by unbalanced training sets in constructing the optimal classifications, this may still introduce problems. In this case, where non-mergers and mergers are apparently difficult to disentangle with the morphology encodings we have chosen, achieving a high completeness comes at the expense of a low purity or PPV, where Figure~\ref{fig:rocevolve} shows $PPV \sim 0.3$ at $z=1.5$ rising to $PPV \sim 0.6$ at $z=3$.

Because the completeness, TPR, values are relatively stable over time, this evolution in PPV could indicate a troubling circular argument:  the intrinsic merger rate in \illustris\ rises steeply over this cosmic time, causing any poor classifier to have an increasing PPV value, and via Equation~\ref{eq:mergerfrac} to imply an increasing merger fraction or rate. There are two reasons to rule out this concern.  First, the rise in merger fraction seen in Figure~\ref{fig:fractions} is significantly greater (factor of 10) than the rise in PPV shown by Figure~\ref{fig:rocevolve}. Therefore, the fact that the RFs infer a rising merger rate (against simulations and data test sets) is not completely a coincidental side-effect of the unbalanced training set, and so we conclude the RFs do provide some value.

Moreover, to concretely determine the effect of an unbalanced training set, we repeat the analyses of this paper using a different formulation for training the RFs. Instead of using the intrinsic \illustris\ galaxy sample, we create an alternative training set that has the same number of mergers and non-mergers. We achieve this by randomly undersampling the (more numerous) class of non-mergers to have the same sample size as the set of mergers. We show RF results with this alternative training set in Figure~\ref{fig:newtraining}. 

This has two main effects. First, the training sets have reduced sizes, and so the RF training prescription leads to somewhat overtraining the alternative RFs. We mitigate this by adjusting down the maximum number of leaf nodes in each forest, which achieves a level of overtraining (still non-zero) similar to the original RFs, by reducing the possible complexity of the RFs. The left panel of Figure~\ref{fig:newtraining} shows that the ROC at $z=1.5$ is very similar to Figure~\ref{fig:roc} and thus the change in training set does not drastically affect the overall performance obtained by the RFs.

The second main effect of this balanced training set is to affect the purity metrics such as PPV, because there are now many fewer potential false positive sources. By construction, the RFs can only select similar regions of morphology space (because the merger sample is the same), and therefore the completeness (TPR) values are similar. Moreover, it does not change the false positive rate FPR, because that metric does not depend on the relative size of the classes in the training set. However, because of the fewer non-mergers relative to mergers, with a similar classification scheme, the purity (PPV) values are much greater in the center panel of Figure~\ref{fig:newtraining}: $PPV \approx 0.8$. Therefore these values are more appropriate to use when contrasting to different classification schemes created using a training set with balanced class sizes.

Finally, we apply the alternatively trained RFs to the analysis of CANDELS and \illustris\ merger fractions in the right panel of Figure~\ref{fig:newtraining}. We find similar results except for $z=0.5$, which is a substantial outlier here. Although the PPV/TPR correction (Equation~\ref{eq:mergerfrac}) is expectedly much smaller, we recover essentialy the same steeply rising merger fractions, albeit again a factor of 2 greater than predicted by theory. Because we achieve essentially the same conclusions with either training set, we confirm that the conclusions based on these RFs are reasonably robust to the construction of the training set.

\begin{figure*}
\begin{center}
\includegraphics[width=2.3in]{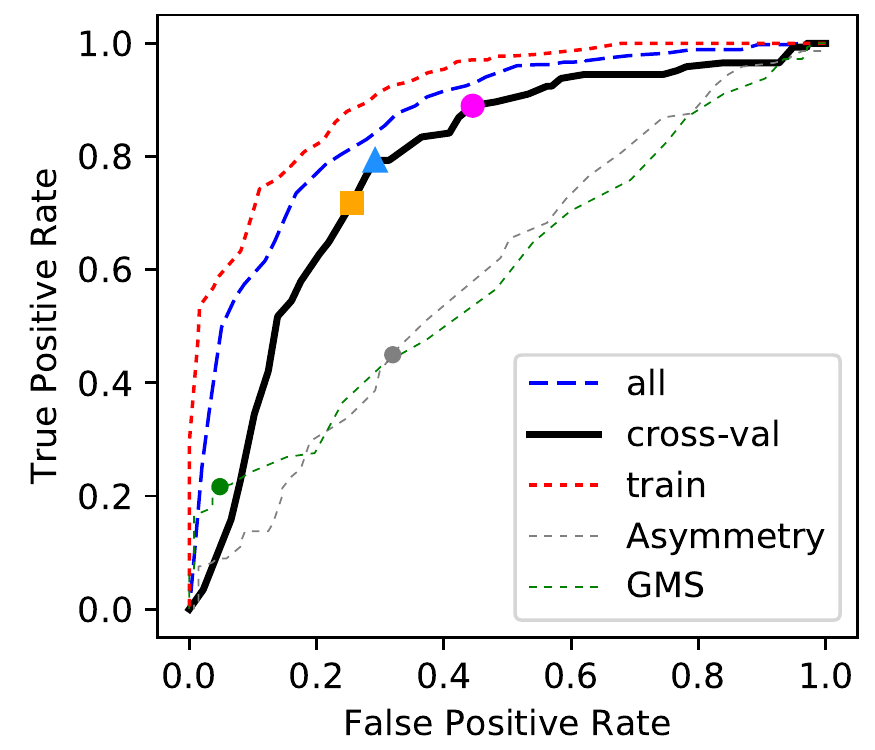} 
\includegraphics[width=2.3in]{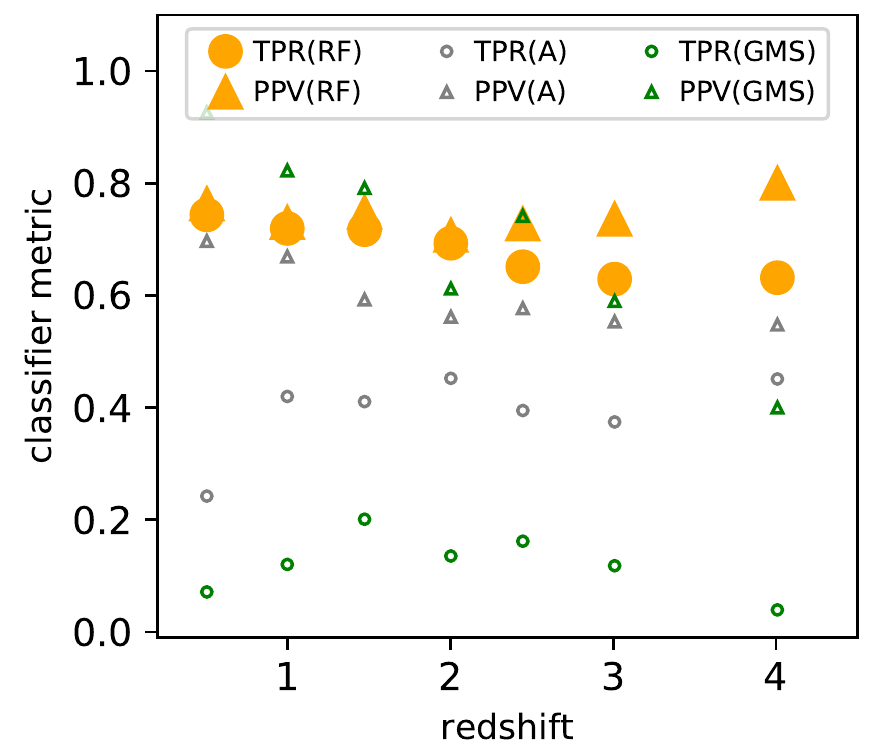} 
\includegraphics[width=2.3in]{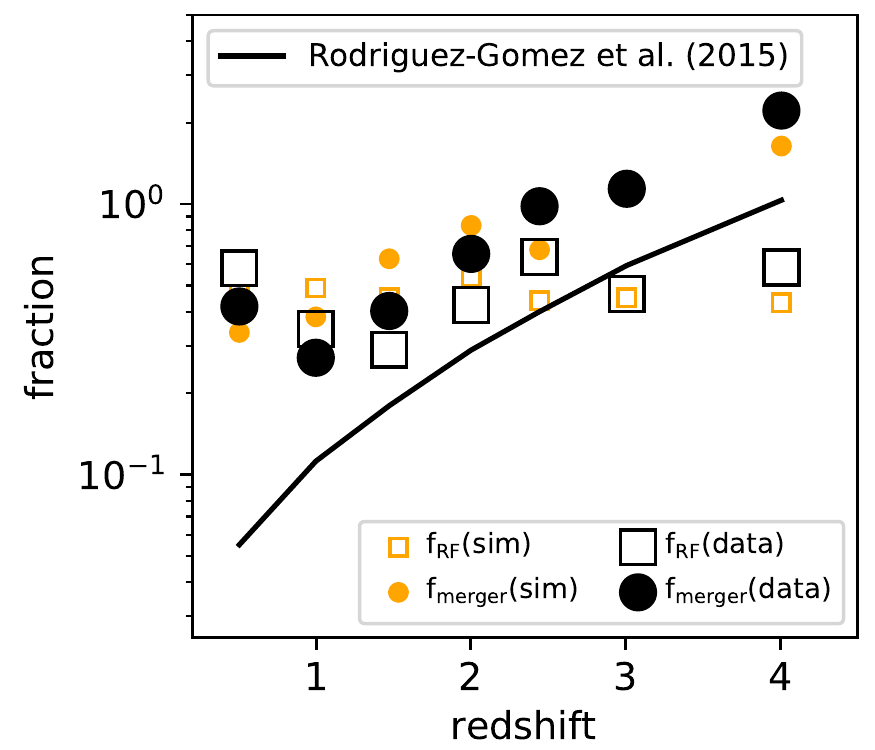}
\caption{  RF results when using random undersampling to create a balanced training set where the number of non-mergers equals the number of mergers. This shows panels identical to those shown in Figures~\ref{fig:roc}, \ref{fig:rocevolve}, and \ref{fig:fractions}. The main difference is that because the training sets have balanced class sizes and the merger class is unchanged, the positive predictive values (PPVs) are much larger than for the default case from Figure~\ref{fig:rocevolve}. This shrinks the size of the sample selected by the RFs in the right panel, but because the PPV/TPR correction is also much smaller, the resulting merger fraction evolution is similar, except for the outlier at $z=0.5$. Therefore the RF-based conclusions are largely insensitive to whether the training set classes are balanced or unbalanced. \label{fig:newtraining}} 
\end{center}
\end{figure*}

}

\begin{figure}
\begin{center}
\includegraphics[width=3in]{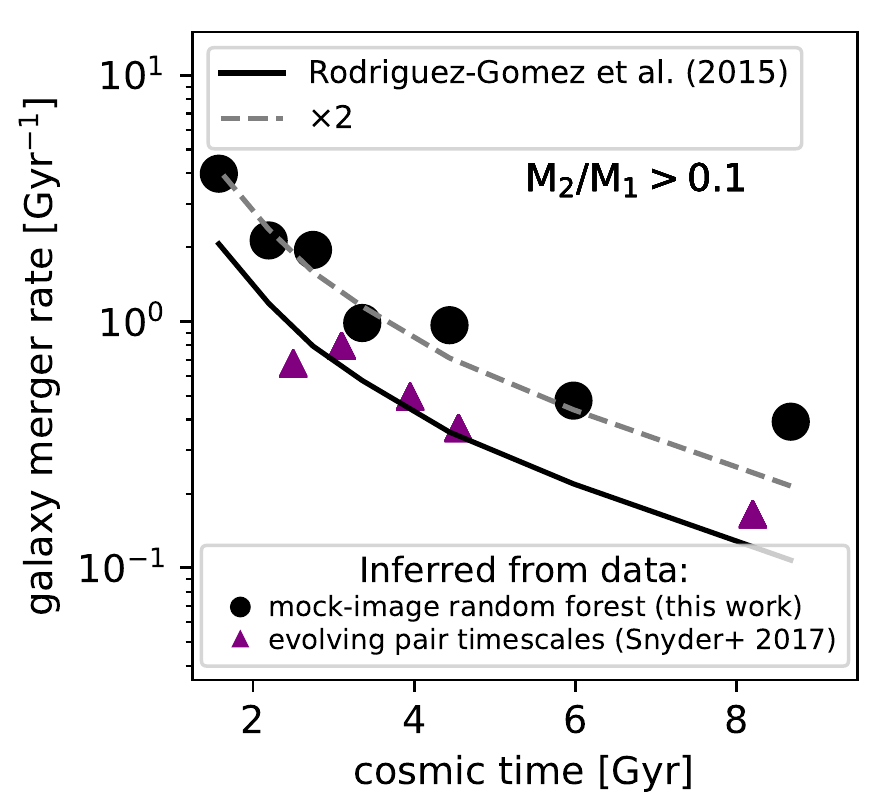}
\includegraphics[width=3in]{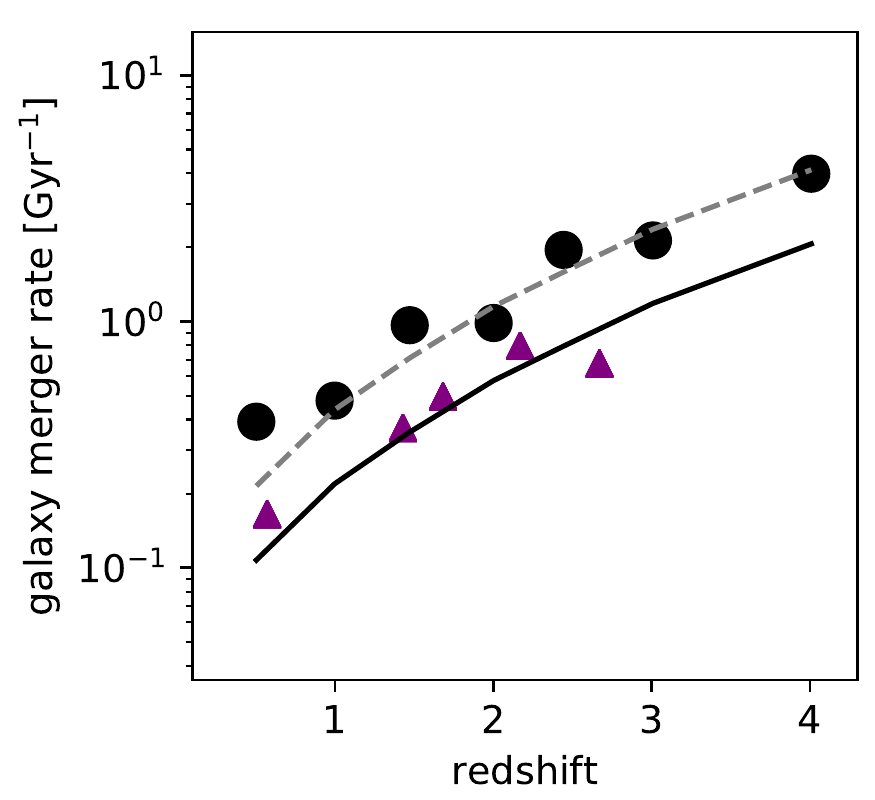}
\caption{Estimated merger rates versus cosmic time and redshift, showing that observations and theory agree that the incidence of galaxy mergers decreases rapidly with time.  The \illustris\ RF-based merger rates are about of factor of 2 larger than inferred from pair statistics, possibly indicating a difference in morphology distribution between simulation and data.  \label{fig:rates}}
\end{center}
\end{figure}

\section{Discussion} \label{s:discussion}

In order to learn how to observe the merging process, we rely on simulations to determine the observability of mergers, in a statistical sense.  By observability, here we mean the strength and duration of enhanced morphological features, which cannot be predicted analytically given the complexity of the physics throughout the merging process. Any set of merger candidates must be calibrated with simulations, in order to know the observability times and thus translate the measured characteristics into a characterization of the underlying physical event(s).

However, all such simulations have flaws, and they are not guaranteed to truly reflect reality. Therefore the translation between observed and intrinsic merger events might lead to incorrect conclusions. However, with the rapid growth of comprehensive cosmological hydrodynamic simulations of large galaxy populations, we can remove some of this ambiguity, for example by naturally capturing effects driven by morphological diversity, the distribution of merger orbits, and cosmological evolution in the relevant timescales \citep[e.g.,][]{Snyder2017}. While any given simulation may not provide a perfect training set, by testing new competing models for interpreting observed processes, we are likely to gain some insight and advance our ability to interpret observations of galaxy assembly over cosmic time. And while in this approach, any given galaxy cannot be classified with complete certainty, we can learn important lessons, as we have found in this paper.

\subsection{Distant Galaxy Merger Rates}

We have shown that the late-stage merger rate evolution implied by our simulation-based RF classification is consistent with pair-based estimates. Figure~\ref{fig:rates} compares the merger rate we infer using the \illustris-based RF classifications of CANDELS images to that inferred from analyzing close pair fractions. We find agreement in the general trend that, assuming evolving pair observability timescales \citep{Snyder2017}, observed pair counts imply a steeply rising merger rate to $z > 2$ \citep[e.g.,][]{Man2016,Ventou2017,Mantha2018}. These observed merger rate trends now agree well with expectations from theory up to $z\sim 3$. 

The RF-based observed merger rates are a factor of $\approx 2$ greater than the rates implied by pair counts (observed and predicted), at the same masses, mass ratios, and redshifts.  This offset could be caused by several factors, such as differences in the morphology distribution between \illustris\ and the real Universe. 

\subsection{Bulges as a Merger Diagnostic?}

Figure~\ref{fig:pastfuture} shows that morphology features typically associated with bulges are important for training the RFs to select mergers completing within the $250$ Myr prior to observation.  This result is noteworthy and it warrants further study.  For instance, this could result from a difference in the intrinsic merger rates of bulge-dominated versus disk dominated galaxies in our samples, plausibly caused by bulge-dominated galaxies residing in more massive dark matter halos {(and thus having a higher merger rate)}.  If so, then the bulge is a secondary indicator of a high probability of finding a merger within the $500$ Myr window and not a direct signature of the merger.  

On the other hand, it is possible that the merger process actually formed the bulge or central concentration in question, in which case these features are direct signposts of recent mergers. This possibility is especially intriguing at higher redshifts, where there is less time for processes to form bulges, in which case the existence of one becomes an even more certain merger tracer.  Intriguingly, we find that the purity of the RF classifications is low at $z=0.5$ and increases to $z=4$.  If mergers form the majority of bulges, and those structures persist, then this is exactly what we would expect to find: the declining purity as time advances results from past mergers that formed bulges outside of the $500$ Myr window, which becomes a smaller fraction of the Universe's age over time. { Furthermore, Figure~\ref{fig:pastfuture} shows that the bulge-merger correspondence is less apparent in the ``future merger'' case. If the bulge strength were simply a proxy for the galaxy mass, i.e. if the link between halo mass and merger rate were causing the coincidental correlation between bulge features and merger labels, then we would have expected $F(G, M_{20})$ and $C$ would have a high importance for future mergers as well. }

Finally, if there is a mismatch in the real versus simulated galaxy morphology distribution driven by the uncertain feedback physics used in \illustris, noting that \illustris\ generally forms too many disc-dominated galaxies, then the RFs might be overly sensitive to bulges. Similarly, if \illustris\ lacks a physical process that forms real bulges, such as disc fragmentation \citep[e.g.,][]{Porter2014}, then the RFs may overestimate the incidence of mergers.

\subsection{Further Improvements}

Although we have constructed merger classifications that improve upon simple one- or two-dimensional schemes, they are far from perfect, yielding samples with up to only $\sim 70\%$ completeness and $\sim 30\%$ purity { when tested against samples with realistic merger fractions. In this case, we found that the RF performances are optimal around $z=1.5$, with purity dropping steeply at $z < 1$ and completeness dropping steeply at $z > 2$. }

In this paper, we used only five feature measurements per wavelength, but there is no guarantee that any five such features will be sufficient for a given similar training set.  Moreover, we used a rigid merger definition that may not be the optimal one for selecting the most interesting range of events.

We are also limited by the size of the training set. Large cosmological simulations help to build sizable training sets, but mergers are still intrinsically rare, and so we trained the RFs with only a few hundred merger events at any given timestep.  Some steps to improve the situation are to use still larger simulations, and to analyze every timestep, where here we considered a pre-existing set of images at only about 10 timesteps.  This would also help to span a wider range of merger characteristics.

In this paper, we used manual encoding of image features to train RF classifications. An alternative approach would use an auto-encoding technique, such as convolutional neural networks, to determine the image features important for selecting mergers \citep[e.g.,][]{Huertas-Company2018}.  In truth, it is likely important to consider both types of investigations: manually encoded features enable us to exploit a long history of intuition of the associated morphology statistics, while auto-encoding techniques may teach us about the rarer, subtler signatures of the merger process.

\section{Summary} \label{s:conclusion}

We created synthetic \hst\ and \jwst\ images from the \illustris\ cosmological simulation, from which we measured common morphological diagnostics.  By combining those diagnostics with knowledge of the intrinsic merger events, we constructed training sets at various epochs ($z=0.5,1,1.5,2,2.5,3,4$).  We used these samples as inputs to ensemble learning techniques, specifically random forests {(RFs)}, to create optimized multidimensional merger classification schemes. We then applied these schemes to existing measurements from the CANDELS Multi-Cycle Treasury program with \hst. We find the following:
\begin{enumerate}
\item{The RFs achieve superior classification results compared with just one- or two-dimensional classifications based on the input morphology statistics such as $A$, $G$, $M_{20}$. Cross-validation shows the RFs yield $\approx 70\%$ completeness, roughly twice that achieved with one- or two- dimensions, with similar sample purity.}
\item{The RFs utilize morphological signatures of mergers occurring throughout the wide time range considered by our training set, $500$ Myr.  Features associated with strong central concentrations or bulges are most important for selecting past mergers (past $250$ Myr), while double nuclei and asymmetries are most important for selecting future mergers (next $250$ Myr). }
\item{When applied to observed surveys, the RFs produce estimated merger rates that rise rapidly from $z=0.5$ to at least $z=3$, confirming complementary probes and agreeing well with theoretical expectations of merger rate evolution.}
\item{The magnitude of the merger rate estimated by the RFs is about twice that implied by theory and other observations, suggesting a possible mismatch in the morphology distribution between \illustris\ and real data. }
\end{enumerate}

\section*{Acknowledgements}
{ We thank the anonymous referee for constructive comments that improved the presentation and content of this work}. GFS thanks Volker Springel, Shy Genel, Chris Hayward, Dylan Nelson, and Joel Primack for discussions about this project. GFS appreciates support from the \hst\ grants program, numbers HST-AR-$12856.01$-A and HST-AR-$13887.04$-A, as well as a Giacconi Fellowship at the Space Telescope Science Institute (STScI), which is operated by the Association of Universities for Research in Astronomy, Inc., under NASA contact NAS 5-26555.  PT acknowledges support provided by NASA through Hubble Fellowship grant HST-HF2-51384.001-A awarded by STScI.  This research has made use of NASA's Astrophysics Data System. Support for programs \#12856 (PI Lotz) and \#13887 (PI Snyder) was provided by NASA through a grant from STScI.  The figures in this paper were constructed with the Matplotlib Python module \citep{Hunter:2007}.  This work used extensively Astropy \citep{Robitaille2013} and NASA's Astrophysics Data System Bibliographic Services.

\bibliographystyle{apj}
\bibliography{$HOME/Dropbox/library}
%\bsp

%%%%%%%%%%%%%%%%%%%%%%%%%%%%%%%%%%%%%%%%%%%%%%%%%%%%%%%%%%%
%%%%%%%%%%%%%%%%%%%%%%%%%%%%%%%%%%%%%%%%%%%%%%%%%%%%%%%%%%%
%%%%%%%%%%%%%%%%%%%%%%%%%%%%%%%%%%%%%%%%%%%%%%%%%%%%%%%%%%%

\end{document}